\documentclass{aa}
\usepackage[varg]{txfonts}
\usepackage{morefloats}
\usepackage{color}
\usepackage{multirow}
\usepackage{graphicx}
\usepackage{fixltx2e}

\newcommand{\hi}{{\sc H\,i}}
\begin{document}
\title{HALOGAS observations of NGC 4414: fountains, interaction, and ram pressure} \author{W.J.G. de Blok\inst{1,2,3} \and G.I.G. J\'ozsa\inst{1}
  \and M. Patterson\inst{4}\thanks{Visiting Astronomer, Kitt Peak National
    Observatory, National Optical Astronomy Observatory, which is
    operated by the Association of Universities for Research in
    Astronomy (AURA) under cooperative agreement with the National
    Science Foundation.} \and G. Gentile\inst{5,6} \and G.H. Heald\inst{1,3} \and
  E. J\"utte\inst{7} \and P. Kamphuis\inst{8,7} \and R.J. Rand\inst{9} \and P. Serra\inst{1,8} \and
  R.A.M. Walterbos\inst{4}{}$^{*}$}
\institute{Netherlands Institute for
  Radio Astronomy (ASTRON), Postbus 2, 7990 AA Dwingeloo, The
  Netherlands 
\and
Astrophysics, Cosmology and Gravity Centre, Department of Astronomy,
University of Cape Town, Private Bag X3, Rondebosch 7701, South Africa
\and
Kapteyn Astronomical Institute, University of Groningen, PO Box 800, 9700 AV, Groningen, 
The Netherlands
\and
Department of Astronomy, New Mexico State University, PO Box 30001, MSC 4500, 
Las Cruces, NM, 88003, USA
\and
Sterrenkundig Observatorium, Universiteit Gent, Krijgslaan 281, 9000, Gent, Belgium
\and
Department of Physics and Astrophysics, Vrije Universiteit Brussel, Pleinlaan 2, 1050, Brussels, Belgium
\and
Astronomisches Institut Ruhr-Universit\"at Bochum, Universit\"atstrasse 150, D-44801 Bochum, Germany
\and
CSIRO Astronomy and Space Science, Australia Telescope National Facility, PO Box 76, Epping, NSW 1710, Australia
\and
Department of Physics and Astronomy, University of New Mexico, 1919 Lomas Blvd NE, Albuquerque, NM 87131-1156, USA
}
\date{Received 21 August 2013 / Accepted 29 April 2014}

\abstract{We present deep \hi\ imaging of the nearby spiral galaxy NGC
  4414, taken as part of the Westerbork HALOGAS (Hydrogen Accretion in
  LOcal GAlaxieS) survey. The observations show that NGC 4414 can be
  characterized by a regularly rotating inner \hi\ disk, and a more
  disturbed outer disk. Modeling of the kinematics shows that the
  outer disk is best described by a U-shaped warp.  Deep optical
  imaging also reveals the presence of a low surface brightness
  stellar shell, indicating a minor interaction with a dwarf galaxy at
  some stage in the past. Modeling of the inner disk suggests that
  about 4 percent of the inner \hi\ is in the form of extra-planar
  gas. Because of the  the disturbed nature of the outer disk, this number is
  difficult to constrain for the galaxy as a whole. These new, deep
  observations of NGC 4414 presented here show that even apparently
  undisturbed galaxies are interacting with their environment.}

\keywords{galaxies: halos -- galaxies: ISM -- galaxies: kinematics and
  dynamics -- galaxies: individual: NGC 4414 -- galaxies: structure}

\titlerunning{NGC 4414}
\maketitle

\section{Introduction}

The last decade has seen a large increase in our knowledge of the
resolved neutral hydrogen distribution in nearby disk galaxies. This
is largely due to the results from surveys such as THINGS
\citep{things}, LVHIS \citep{lvhis}, LITTLE THINGS \citep{hunter},
VLA-ANGST \citep{ott}, FIGGS \citep{figgs}, and SHIELD
\citep{cannon}. These surveys have all yielded detailed information on
the morphology and dynamics of well over a hundred nearby galaxies. An
important science goal of many of these surveys is to explore the
physical processes that link gas and star formation within the galaxy
disks. To probe the relevant size scales (down to those of Giant
Molecular Clouds), these surveys usually focus on maximizing spatial
resolution.

What happens to the gas during other phases of the gas-star formation
cycle is a question these surveys have so far only addressed to a much
smaller extent. How much gas is expelled from the disk due to star
formation (``galactic fountain''; \citealt{shapiro}); how much gas
stays in the halo and how much eventually falls back on the disk?  Is
all the cold gas in the halo due to star formation processes? Can we
also find gas accreted from intergalactic space (the ``cosmic web'')?
An overview of these issues is given in \citet{sancisi}.  However, the
opposite processes are also at work.  Just as gas can be accreted, it
can also be lost again, usually through interactions or stripping due
to the environment.

As noted before, many of the current large \hi\ surveys of nearby
galaxies are optimized for resolution, and may not always reach the
column density sensitivity needed to probe the gas accretion and gas
loss at significant levels.  The Westerbork HALOGAS (Hydrogen
Accretion in LOcal GAlaxieS) survey \citep{heald11} was designed to
overcome this problem.  It used the Westerbork Synthesis Radio
Telescope (WSRT) to observe a sample of $\sim 20$ disk galaxies in the
local Universe (out to 11 Mpc, as based on the distances given in
\citealt{t88}) for 10 times longer than ``typical'' \hi\ observations
and has produced some of the most sensitive interferometric \hi\ data
available. The survey is sensitive to angular scales which are most
useful for studying faint, diffuse gas in and around nearby galaxies,
and has the primary goal of revealing the global characteristics of
low column density gas in and around spiral galaxies in the local
universe.

As described in \citet{heald11}, the HALOGAS sample is selected on the
basis of a number of objective criteria, mainly Hubble type, size,
distance, and inclination. This leads to a sample containing a great
variety of global properties, and, in particular, \hi\  morphologies.

\subsection{NGC 4414}

Here we present an analysis of NGC 4414. Global properties of NGC 4414
as taken from \citet{heald11,heald12} are listed in Table
\ref{tab:summ}.  We refer to these papers for more information.  The
distance of NGC 4414 as determined by the HST Hubble Key Project is
17.8 Mpc \citep{hstkey}, and this is also the distance adopted
here. This distance is therefore significantly different from the 9.7
Mpc given in \citet{t88} (cf.\ Table 1 in \citealt{heald11}).  The
optical appearance of NGC 4414 is characterized by a flocculent spiral
structure. It has the highest star formation rate (4.2 $M_{\odot}$
yr$^{-1}$) of the galaxies in the HALOGAS sample \citep{heald12, h2,
  i4}. This rate is, however, not extraordinarily high for a spiral
galaxy, and NGC 4414 certainly cannot be said to be undergoing a
starburst.  The galaxy has a prominent, smooth, and regularly rotating
molecular gas disk which coincides with the main star forming body
\citep{sakamoto, braine97}. Previous \hi\ observations
\citep{braine93, thornley97} showed a gas disk extending far beyond
the optical disk.  The inner part of this neutral gas disk was found
to be in regular rotation, but the outer parts showed signs of
asymmetries. The emphasis in these previous studies was on the link
between gas, spiral structure, and star formation.  Two other studies
concentrated on the mass-to-light ratio of the stellar disk
\citep{vallejo02} and the dark matter distribution
\citep{vallejo03}. These show that the inner disk of NGC 4414 is
dominated by the stellar population, with only a small dark matter
contribution.

\begin{table}
\caption{Properties of NGC 4414}
\label{tab:summ}
\centering
\begin{tabular}{l l}
\hline\hline
Property   & Value \\
\hline
type              & SAc \\
distance  $D$        & 17.8 Mpc \\
systemic velocity $V_{
\rm sys}$ & 720 km s$^{-1}$ \\
inclination $i$      & 50$^{\circ}$ \\
optical diameter $D_{25}$ & 4.5$'$ \\
absolute magnitude $M_B$ & $-19.12$ \\
rotation velocity $V_{
\rm rot}$ & 224.7 km s$^{-1}$ \\
star formation rate & 4.2 $M_{\odot}$ yr$^{-1}$ \\
\hline
\end{tabular}
\tablefoot{Source: \citet{heald11,heald12}}
\end{table}

%\begin{deluxetable}{llcrccrl}
%\tabletypesize{\scriptsize}
% \tablecaption{Properties of template rotation curves.\label{tab:template}}
% \tablewidth{0pt}
%\tablehead{
%\colhead{$t$} & \colhead{$M_I$}  & \colhead{$\Delta M_I$} & \colhead{$V_0$} & \colhead{$r_{PE}/h$} & \colhead{$\alpha$} & \colhead{$h$} & \colhead{$V_{\rm max}$} \\
%\colhead{(1)} & \colhead{(2)} & \colhead{(3)} & \colhead{(4)} & \colhead{(5)} & \colhead{(6)} & \colhead{(7)} & \colhead{(8)} 
%} 
%\startdata
%0 &  $-23.8$ &  0.4          & 270   & 0.37    & 0.007    & 10.4  & 294 \\
%1 &  $-23.4$ &  0.4          & 248   & 0.40     & 0.006    & 8.6 & 267 \\
%9 &  $-19.0$ &  2.0          & 64   & 0.72     & 0.087    & 1.1  & 102 
%\enddata
%\end{deluxetable}

The consensus in these papers is that NGC 4414 has no close
companions, has not recently suffered a major interaction, but also
that there is some evidence of weak interactions having taken
place. The observed asymmetry of the outer \hi\ distribution suggests
this, but it is also supported by asymmetries in the magnetic field
\citep{soida}.

The galaxy NGC 4414 is part of the Coma I group.  
%Several galaxies in this group
%are found to be \hi-deficient with the highest \hi-deficiency (by
%about a factor of four) in the central $\sim 0.7$ Mpc diameter central
%region (corrected to our assumed distance) \citep{garcia}. The
%galaxies outside this region are either slightly deficient (like NGC
%4414, which has a projected distance of $\sim 0.6$ Mpc from the
%centre) or have normal \hi\ contents. The \hi\ deficiency suggests
%that processes such as ram pressure stripping of the \hi\ disks are at
%work in this group. 
Other HALOGAS targets, such as NGC 4062, NGC 4274, NGC 4448, NGC
  4565, and NGC 4559 are also found in or near the Coma I
  group. Together, these observations can be used to study the
  effects on the outer parts of their \hi\ disks of environment and
  location within the group, however, in this paper the focus is on NGC 4414.
% Part of the focus of this paper will therefore also be on the
%distribution of the gas in the outer parts of NGC 4414 and the
%information it can provide regarding the interaction with the
%intra-group medium.

The paper is organized as follows. In Sect.\ 2 we describe the
\hi\ and optical data. Section 3 contains the kinematical modeling of
the \hi\ disk while in Sect.\ 4 we present an interpretation of the
models. In Sect.\ 5 we give a brief summary.

\section{Data}\label{sec:method} 

\subsection{\hi\  data}

As described in \citet{heald11}, the data were taken using the
WSRT. We also refer to that paper for a description of the
observational set-up and data reduction. For each galaxy in the
HALOGAS sample, two standard data cubes were created.  One is created
using a robust parameter of 0 for intermediate resolution and
sensitivity. To maximize sensitivity to faint extended emission a
second data cube is produced using a Gaussian $u,v$ taper
corresponding to 30$''$ in the image plane. For the WSRT array,
  this taper gives the best compromise between resolution and column
  density sensitivity (cf.\ description in \citealt{heald11}). We
will use both data cubes in this paper, but the analysis of the
dynamics will lean heavily on the 30$''$ tapered cube.

The synthesized beam size of the tapered cube is $39.0'' \times
33.5''$, with a channel spacing of 4.12 km s$^{-1}$. HALOGAS cubes are
Hanning-smoothed, and the channel spacing equals the velocity
resolution. The $1\sigma$ noise per channel in the tapered cube is
0.23 mJy beam$^{-1}$, resulting in a $1\sigma$, 1-channel column
density sensitivity of $0.8 \cdot 10^{18}$ cm$^{-2}$ or, more typical
for an \hi\  profile, a sensitivity of $1.0 \cdot 10^{19}$ cm$^{-2}$ for
a $3\sigma$, 16 km s$^{-1}$ signal.

The robust-weighted cube has a beam size of $25.9'' \times
13.7''$. Its noise per channel is 0.19 mJy beam$^{-1}$, resulting in a
$1\sigma$, 1-channel column density sensitivity of $2.4 \cdot 10^{18}$
cm$^{-2}$ or $2.2\cdot 10^{19}$ cm$^{-2}$ for a $3\sigma$, 16 km
s$^{-1}$ \hi\  profile.

Moment maps were created by first convolving the tapered cube to
$60''$, and clipping the resulting smoothed cube at the $2.5\sigma$
level. The clipped, smoothed cube was used as a mask for the original
resolution cube.  The moment maps were created from the masked,
original cube, with the added constraint that at each position
emission had to be present in a minimum of three consecutive channels
in the smoothed cube.  Lastly, and only for cosmetic reasons, a few
remaining spurious noise peaks were removed. The same mask and
procedure was used for the robust cube.

In order to keep uniform noise properties we do not make a correction
for the primary beam. The only exception is when measuring total
fluxes. However, we found the corrections there to be $\sim 1\%$ and
therefore negligible compared to other uncertainties.

While the noise in a channel, $\sigma_{\rm chan}$, and therefore the
sensitivity in a channel, is well-defined, the sensitivity in an
integrated column density map (zeroth moment) depends on $\sigma_{\rm
  chan}$ as well as on the number of channels $N$ that are
contributing to each spatial position in the map. For a
Hanning-smoothed cube with independent channels the zeroth moment map
noise $\sigma_{\rm mom}$ is defined as $\sigma_{\rm mom} = \sqrt{N}
\sigma_{\rm chan}$. We created a map of $\sigma_{\rm mom}$ and by
comparing this with the integrated \hi\ map itself, determined the
signal-to-noise $S/N$ in the integrated \hi\ map as a function of
position. The average integrated intensity value of all pixels with
$4.5 < S/N < 5.5$ is 0.015 Jy beam$^{-1}$ km s$^{-1}$ for the tapered
cube and 0.016 Jy beam$^{-1}$ km s$^{-1}$ for the robust cube. Because
of the clipping used before we do not expect $\sigma_{\rm mom}$ to
have a Gaussian distribution. Nevertheless we can use the number
derived here as an estimate for the column density sensitivity,
leading to $S/N \sim 5$ column density values of $5.3 \cdot 10^{19}$
cm$^{-2}$ for the tapered cube and $2.1 \cdot 10^{20}$ cm$^{-2}$ for
the robust cube (this is an averaged number derived from all pixels
with $4.5 < S/N < 5.5$, as described above).  These numbers differ
from the column density sensitivity given earlier.  The latter refer
to the sensitivity per channel, whereas in a zeroth-moment map
multiple channels contribute to one pixel, increasing the noise in the
zeroth-moment map at that position. (The noise increases as
$\sqrt{N}$ while the signal increases with $N$; isolating regions with
signal thus means an effective $\sqrt{N}$ increase in sensitivity in
the moment map).

\begin{figure*}
\includegraphics[width=0.9\hsize]{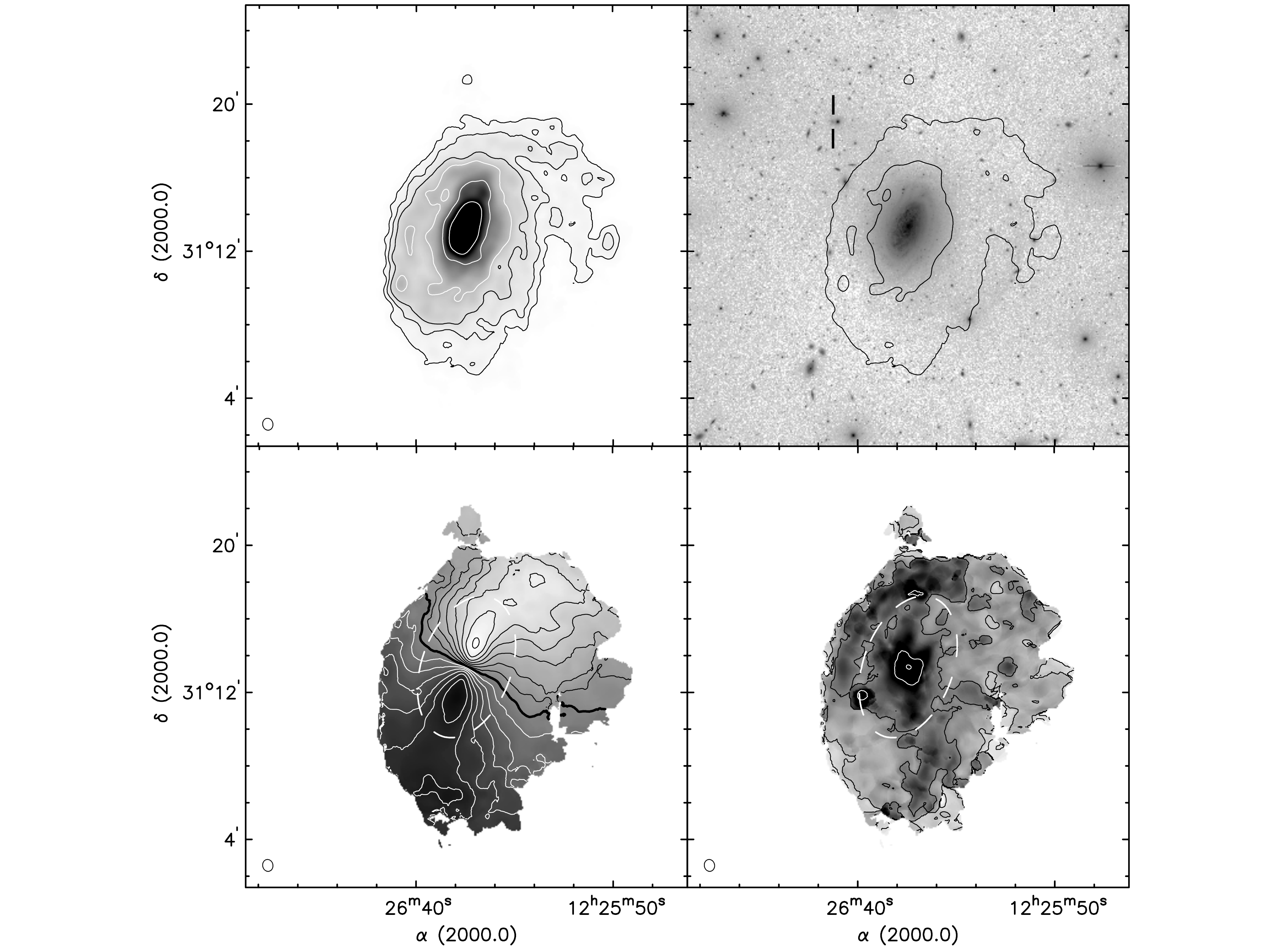}
\caption{Overview of NGC 4414 derived from the $30''$ tapered data
  cube. \emph{Top-left:} integrated \hi\ intensity map (zeroth
  moment). The contour levels shown are $(2, 5, 10) \cdot 10^{19}$
  cm$^{-2}$ (black) and $(2, 5, 10) \cdot 10^{20}$ cm$^{-2}$
  (white). \emph{Top-right:} Integrated \hi\ intensity overplotted on
  a deep KPNO $B$-band image. The optical image is shown using a
  logarithmic intensity scale.  Contours shown are $(2, 20) \cdot
  10^{19}$ cm$^{-2}$. The two short vertical lines indicate galaxy
  SDSS J122646.27+311904.8, as discussed in
  Sect.\ 4.2. \emph{Bottom-left}: Intensity-weighted mean first-moment
  velocity field. The thick black countour shows the systemic velocity
  of 711.5 km s$^{-1}$. Black contours then decrease in steps of 25 km
  s$^{-1}$, white contours increase in steps of 25 km s$^{-1}$. The
  dashed white ellipse indicates the division between inner and outer
  disk as defined in Sect.\ 3. \emph{Bottom-right:} Second-moment
  velocity dispersion map. Black contours show values of (5, 10, 20)
  km s$^{-1}$ (light-gray to dark-gray background). White contours
  show (50, 100) km s$^{-1}$ (dark-gray to black background). The
  white ellipse indicates the division between inner and outer
  disk.\label{fig:moms}}
\end{figure*}

The moment maps and velocity field derived from the tapered cube are
shown in Fig.~\ref{fig:moms}. We see that the \hi\  distribution of NGC
4414 is lopsided, with the degree of lopsidedness depending on the
column density level. At intermediate column densities, the disk
extends farther towards the SE. In contrast, at lower column densities
the disk column densities drop much faster is towards the SE than to the NW.
This NW part of the disk breaks up into what appear to be fragments of
arms and clumps.  The inner part of the velocity field is clearly
dominated by rotation, and symmetrical, despite the disturbed looking
\hi\  morphology. The southern part of the disk seems to show regular
rotation out to large radii, whereas the northern part looks more
disturbed with a large change in kinematical PA.  The extreme outer
eastern and western edges of the disk also show strong kinks in the
velocity contours indicating streaming motions or strong
discontinuities in PA or inclination.

The second-moment map (indicative of the velocity dispersion for
symmetrical, single Gaussian profiles) shows a significant area with
values higher than 20 km s$^{-1}$. Most of this is, however, is not
caused by an intrinsically high velocity dispersion. The central area
with high dispersion is due to beamsmearing caused by the steep
rotation curve.  The curved area with high values to the NE is due to
double and asymmetric profiles where a separate \hi\ feature is seen
projected onto the main disk (this coincides with the strong
kinks in the velocity field contours).  Also shown in the Figure is an
overlay of the \hi\ emission on top of an optical $B$-band image
(described below in Sect.\ 2.2).

\begin{figure*}
\includegraphics[width=0.9\hsize]{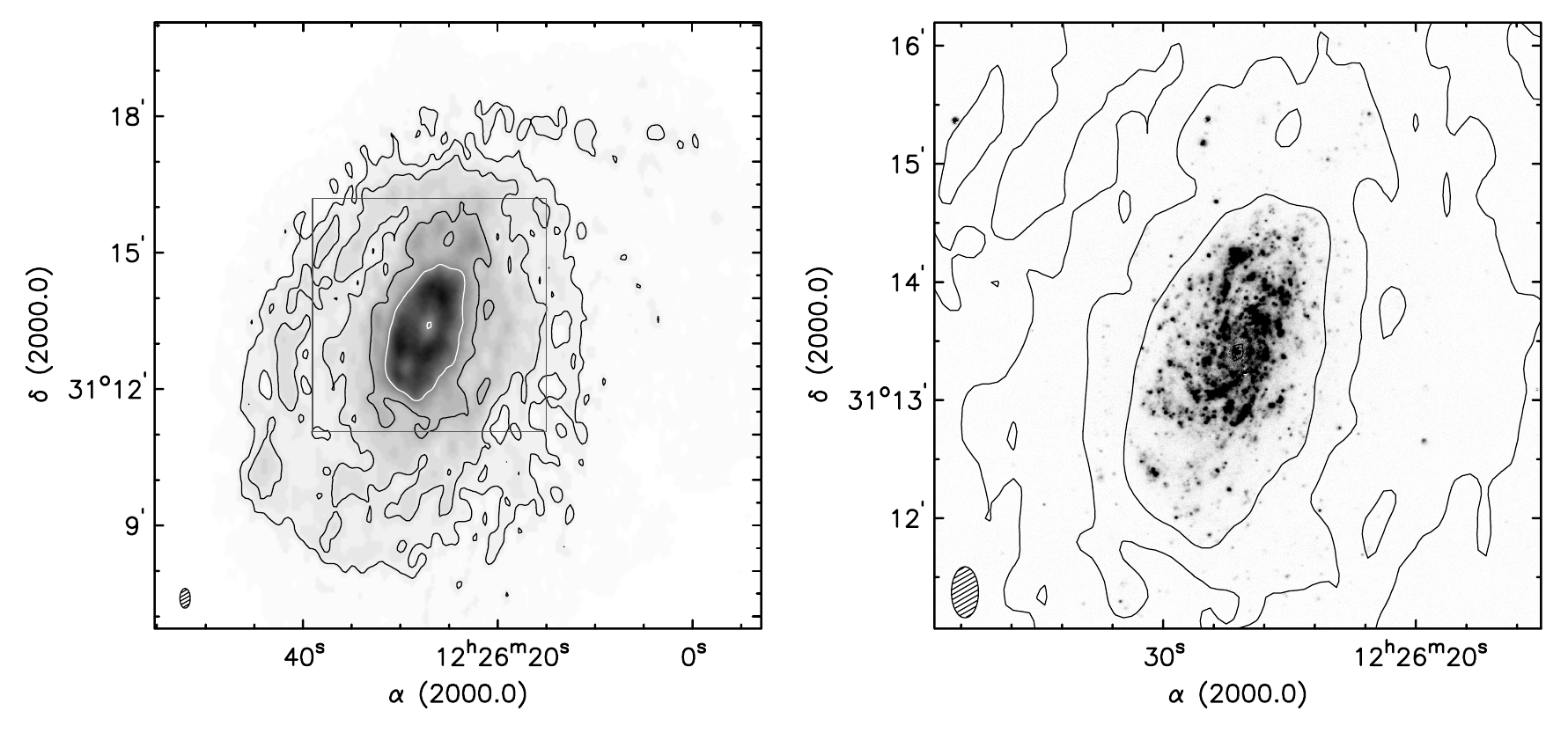}
\caption{\emph{Left:} Robust-weighted integrated column density map.
  Contour levels are $(1, 2, 5) \cdot 10^{20}$ cm$^{-2}$ (black) and
  $1 \cdot 10^{21}$ cm$^{-2}$ (white). The gray box shows the area
  shown in the right-panel. \emph{Right:} Integrated \hi\  column density
  based on the robust weighted cube overplotted on a deep KPNO H$\alpha$
  image. Contour levels are the same as in the left panel. The main H$\alpha$ distribution is encompassed by the $1 \cdot
  10^{21}$ cm$^{-2}$ contour, with low-level H$\alpha$ primarily found
  within the $5 \cdot 10^{20}$ cm$^{-2}$ contour. Virtually no
  H$\alpha$ is found outside these countours. The bright source in the
  top-left is a residual continuum artifact.\label{fig:robhalfa}}
\end{figure*}

An integrated column density map based on the robust cube is shown in
Fig.\ \ref{fig:robhalfa}.  It is clear that the disk is better
resolved but at the cost of lower sensitivity. An overlay of the
robust zeroth moment map on top of an H$\alpha$ image (details below
in Sect.\ 2.2) is shown in the same figure. The high-level
star formation is encompassed by the $1 \cdot 10^{21}$ cm$^{-2}$
contour, with lower-level star formation happening within the $5 \cdot
10^{20}$ cm$^{-2}$ contour.  No additional massive star formation is
detected outside the $2\cdot 10^{20}$ cm$^{-2}$ contour.  The two
distinct levels of star formation in the inner and outer disk are
consistent with the classification of NGC 4414 as a Type 1 XUV disk
(\citealt{thilker}; Type 1 XUV disks are defined as showing structured
UV emission beyond the location of the star formation threshold). In
their description of NGC 4414, they mention the XUV clumps ``hugging
the main disk'', with some of them coinciding with H$\alpha$ clumps.

\begin{figure*}
\centerline{\includegraphics[width=0.9\hsize]{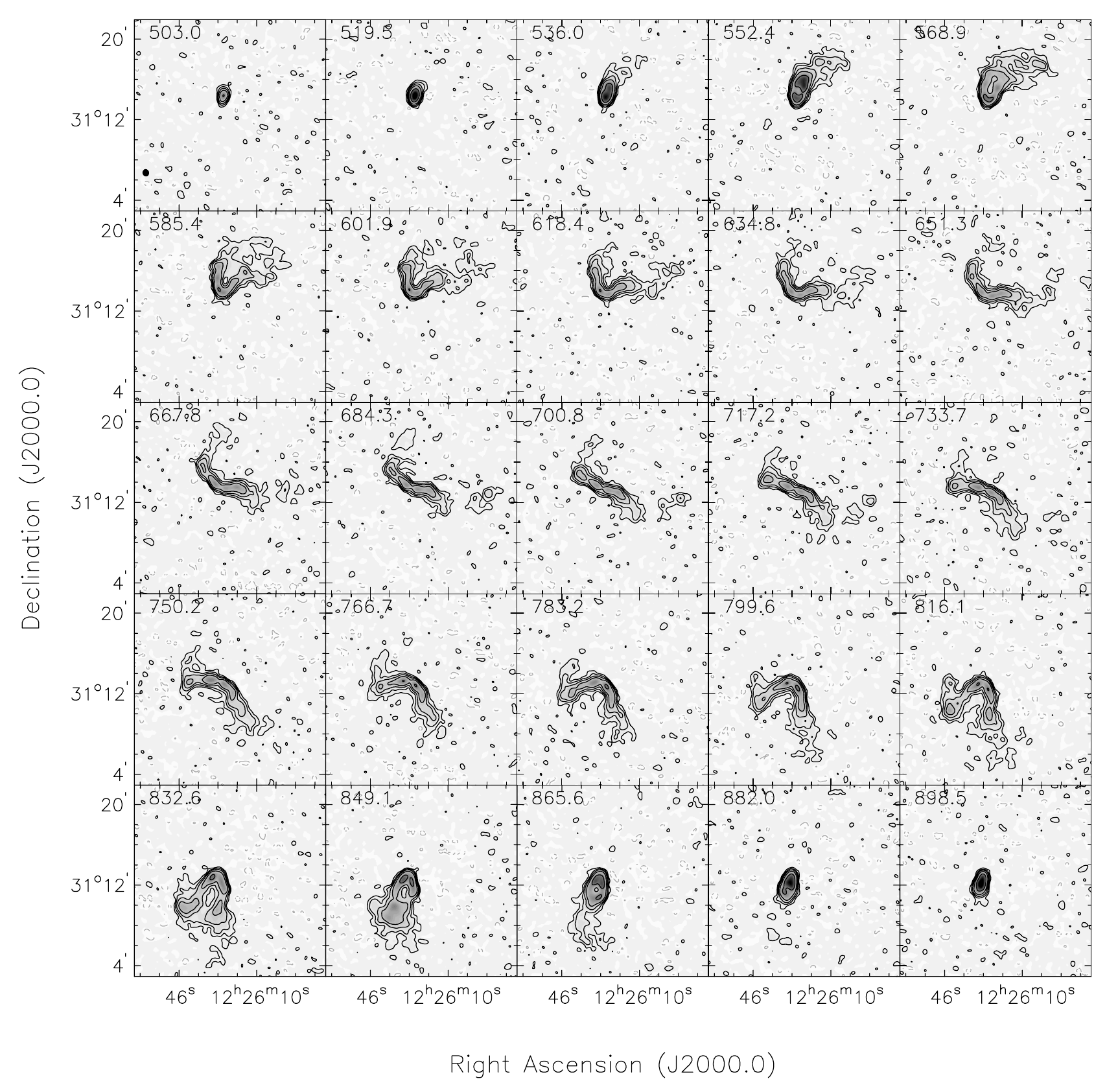}}
\caption{Selected channel maps from the 30'' tapered cube. Only every
  fourth channel map is shown. Contour levels are $(2, 5, 10, 20, 50)
  \cdot \sigma$ where $\sigma=0.23$ mJy beam$^{-1}$ or $0.8 \cdot
  10^{18}$ cm$^{-2}$. The beam is indicated in the bottom-left of the
  top-left panel. The heliocentric velocity in km s$^{-1}$ is
  indicated in the top-left of each panel.\label{fig:chans}}
\end{figure*}

Selected channel maps of the 30$''$-tapered cube are shown in
Fig.\ \ref{fig:chans}. Note the sharp twist in position angle between
inner and outer disk emission (e.g., at $V=568.9$ km
s$^{-1}$). Similar sharp twists remain visible throughout most of the
data cube. The inner, higher column density emission does show the
signatures of regular rotation.

\begin{figure}
\includegraphics[width=0.9\hsize]{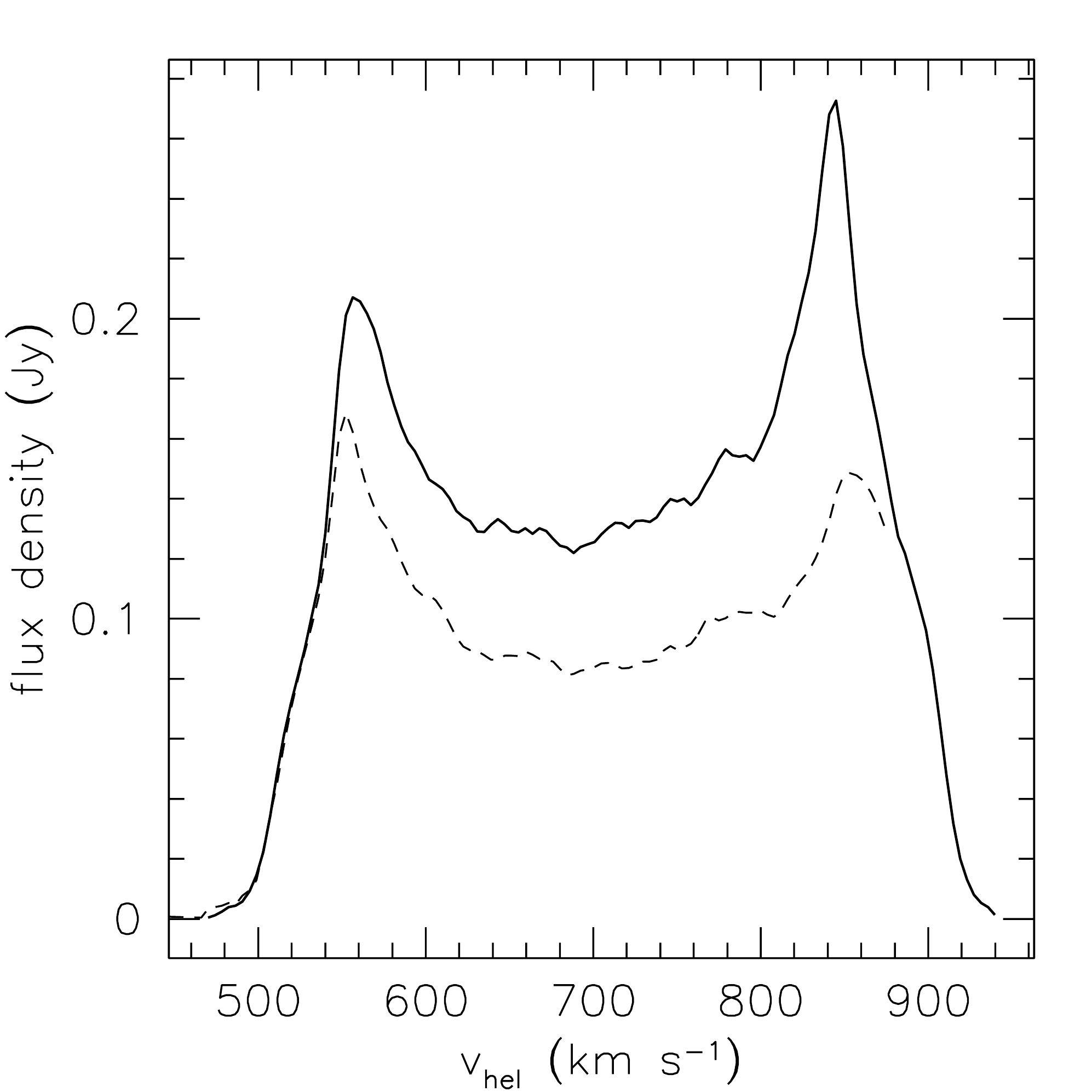}
\caption{Global \hi\  profile based on the primary-beam corrected
  30''-tapered cube (full line). The dashed line shows the global
  profile of the inner disk only.\label{fig:glob}}
\end{figure}

We also derived a global \hi\ profile using a primary-beam corrected
tapered cube. This profile is shown in Fig.\ \ref{fig:glob}. The
profile is asymmetric with one of the horns more prominent than the
other, reflecting the disturbed nature of the outer part of the
galaxy. The total flux derived from the profile is 60.4 Jy km
s$^{-1}$, translating into a total \hi\ mass of $4.5 \cdot
10^{9}\ M_{\odot}$. This corresponds well with previous
determinations. For example, Braine et al.\ (1993) find a total flux
of 64.9 Jy km s$^{-1}$ and an \hi\ mass of $4.8 \cdot
10^{9}\ M_{\odot}$ (corrected to our assumed distance). The small
difference can almost certainly be attributed to the lower sensitivity
of the older observations ($1 \times 12^h$ instead of $10 \times
12^h$) leading to larger susceptibility to the effects of noise. We
measure velocity widths of $W_{20} = 391$ km s$^{-1}$ and $W_{50} =
333$ km s$^{-1}$ (not corrected for channel width).

\subsection{Optical data}

NGC 4414 was observed on March 20, 2012 in $B$, H$\alpha$, and $R$
(for continuum-subtraction) on the Mosaic 1.1 Wide Field Imager
instrument on the KPNO 4-meter telescope.  The weather was clear with
moderate wind with typical seeing of $\sim 1.2''$.  The Mosaic
instrument has eight 1k $\times$ 4k CCDs with 15 micron pixels,
resulting in a field of view of $36'$ on a side and 0.26 arcsec
pixel$^{-1}$.  At the distance of NGC 4414 (17.8 Mpc), the Mosaic
yields a field of view of 186 kpc on a side with 22.4 pc pixel$^{-1}$.
We used the FillGap dither command to observe, which offsets the
telescope with a five-exposure pattern in order to cover the 20-arcsec
gaps between CCDs. We observed NGC 4414 for a total of 30 ($5 \times
6$) minutes in H$\alpha$, 90 ($15 \times 6$) minutes in $B$, and 10
($5 \times 2$) minutes in $R$. The H$\alpha$ filter is centered at
6575 \AA{} with a full-width at half maximum (FWHM) of 81 \AA, which
includes the [NII] emission lines at 6548 \AA{} and 6584 \AA. We
assume a ratio of [NII]/H$\alpha$ of 0.4 for H$\alpha$ flux
measurements. Only the $B$ and H$\alpha$ images are used in this
paper.

For the data reduction, we used the MSCRED Mosaic Data Reduction
Package within the Image Reduction and Analysis Facility (IRAF).  We
used CCDPROC for the basic reduction and implemented additional
recommended steps including a secondary flat field correction and
ghost pupil reflection artifact removal (present in $B$ and H$\alpha$,
but not $R$).

%We first removed crosstalk between CCDs, trimmed the images, applied
%the bias level correction, created bad pixel masks to mask cosmic rays
%and saturated stars, and applied a first flat field correction using
%dome flats taken in each filter. Using the dome flat fielded images,
%we constructed first pass super sky flat fields by masking all stars
%and galaxies in selected small angular size target science images and
%stacking the masked images to produce images of the sky in $B$,
%H$\alpha$, and $R$, which we used to apply a secondary flat field
%correction.  Before applying the sky flat field correction to our $B$
%and H$\alpha$ science images, we first removed the ghost pupil
%reflection artifact by creating a pupil template from the first pass
%super sky flat field and then scaling and subtracting the template
%image from our science images using the IRAF tasks MSCPUPIL and
%IRMPUPIL. After removing the ghost pupil, we constructed a final super
%sky flat field and applied it to our images.

%We applied world coordinate system solutions to each image with
%MSCCMATCH, removed the background sky gradient using MSCSKYSUB, and
%combined the dithered exposures for each filter. To create continuum
%subtracted Ha images, we used MSCIMAGE to determine the scale and
%coordinate offsets of stars in the R band images compared to the same
%stars in the Ha images.

We flux-calibrated our images based on observations of
spectrophotometric stars and standard star observations over a range
of airmasses taken throughout the four night observing run.  The
$B$-band image has a zeropoint of $23.33 \pm 0.03$ mag arcsec$^{-2}$
and an rms of 0.042 counts/sec or 26.78 mag arcsec$^{-2}$.  The final
H$\alpha$ flux-calibrated image has an rms noise of $\sim 3 \times
10^{34}$ ergs s$^{-1}$ at the distance of NGC 4414.  This is deep
enough to detect a faint H\,{\sc ii} region ionized by a single B0
star.

In addition, NGC 4414 was observed in $r'$ band in 2010 February 4, at
the Isaac Newton Telescope (INT), La Palma, Spain. The observations
were conducted as part of the HALOSTARS campaign. The goal of
HALOSTARS is to map HALOGAS galaxies with the wide field camera (WFC)
of the INT to a sufficient depth to detect faint, extended stellar
features that might indicate recent interactions of the target
galaxies. Sky flats and bias frames were obtained at the beginning and
at the end of the night. A total of 17 exposures of $300\,{\rm s}$
each were taken to obtain a total of $5100\,{\rm s}$ on-source
integration time under variable weather conditions in presence of the
moon. The projected size of NGC 4414 is too large to fit on a single
INT WFC chip. We hence used a wide dithering scheme to make a mosaic
of the source. To process the raw data, we made use of the wide-field
imaging reduction pipeline THELI \citep{schirmer13}. After
overscan- and bias-correcting the images, the chips were flatfielded
using the sky flats in addition to a smoothed averaged night sky
calculated from the exposures on NGC 4414 (a ``super-flat''). Image
errors (hot and cold pixels, cosmics) were automatically
masked. Using SCAMP \citep{bertin2006}, a (relative) photometric
calibration was applied, and the images were background-subtracted and
co-added, solving for astrometric distortion by making use of the SDSS
(DR8) catalog. Also using the SDSS catalog, an absolute
calibration was applied. The zeropoint was determined to be $24.67 \pm
0.1$ mag. The rms noise of 0.027 ADU s$^{-1}$ implies a 1$\sigma$
level of 26.2 mag arcsec$^{-1}$.  Both images will be further discussed and
presented in Sect.\ 4.2.

\section{Modeling}

\begin{figure}
\includegraphics[width=0.9\hsize]{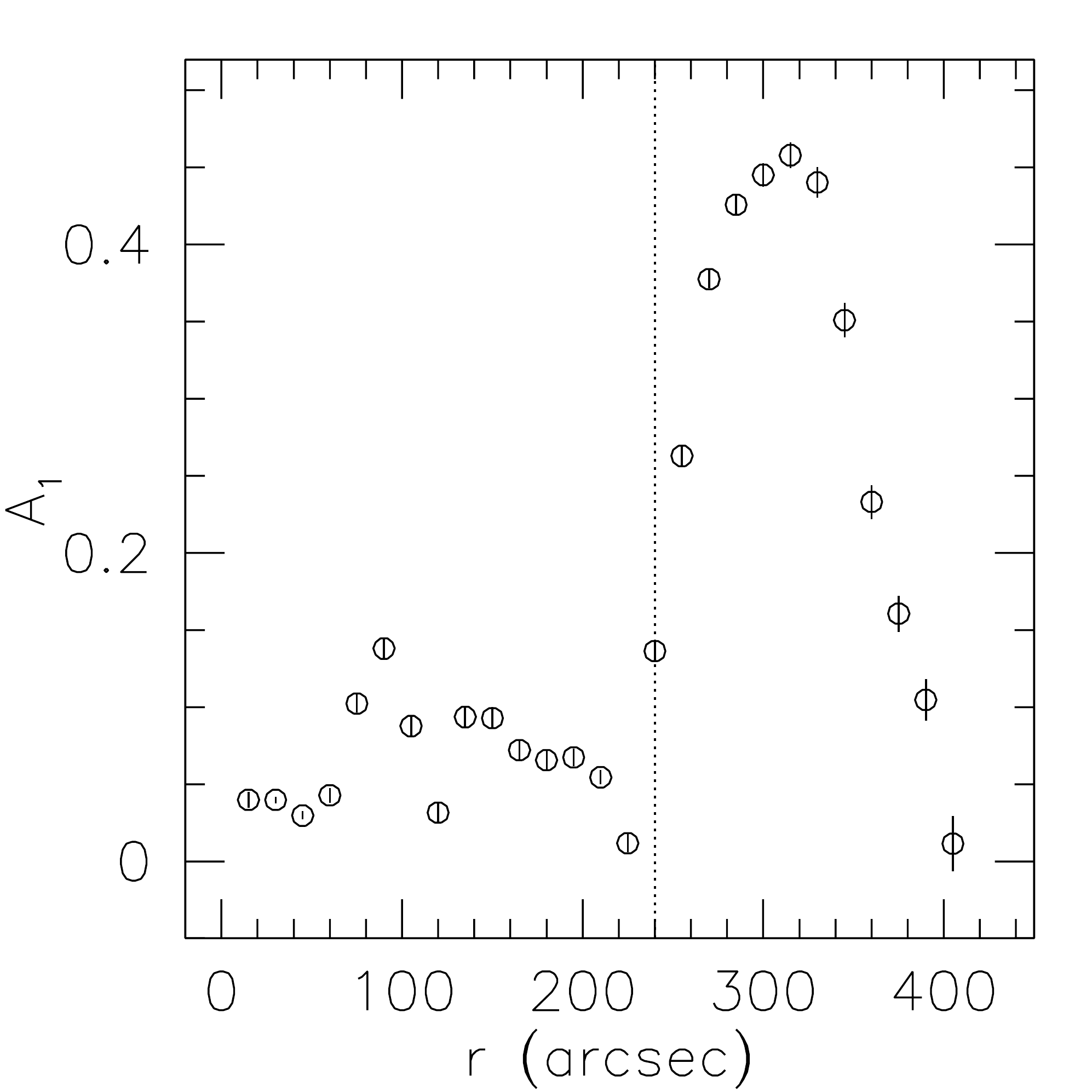}
\caption{Normalized lopsidedness parameter $A_1$ as a function or radius. The different
 behavior in inner and outer disk (separated by the vertical line at a radius of $240''$) is clear.
\label{fig:lop}}
\end{figure}

As noted earlier, in kinematical terms, NGC 4414 can be divided into a
symmetrical, regularly rotating inner disk, and a more asymmetrical,
somewhat disturbed outer disk. The disturbance of the outer disk can
be quantified using a Fourier decomposition of the H\,{\sc i}\ surface
density $\sigma$ as described by \cite{rix95}. We use $\sigma(r,\phi)
= a_0(r) + \sum a_m(r) \cos(m\phi - \phi_m(r))$, where $a_0(r)$ is the
mean surface density, $\phi$ the azimuthal angle in the plane of the
galaxy and $a_m$ and $\phi_m$ the $m^{th}$ amplitude and phase of the
harmonic coefficient. Lopsidedness can be characterized by an $m=1$
mode. Here, we calculate the normalized amplitude $A_1(r) =
a_m(r)/a_0(r)$. Figure \ref{fig:lop} shows the variation of $A_1$ with
radius. We see a distinct difference between inner and outer disk. The
lopsidedness parameter $A_1$ is low (mostly $<0.1$) in the relatively
symmetric and undisturbed inner disk. In the outer disk, it increases
rapidly up to 0.5, which is far above the commonly adopted
lopsidedness threshold ($A_1 = 0.1$--$0.2$; see, e.g.,
\citealt{angiras06,vaneymeren11b,zaritsky13}).

We analyze and model the kinematics of the inner and outer disk
separately. Based on the lopsidedness we define the inner disk as the
part of the galaxy inside a radius of 240$''$. The outer disk consists
of the remaining disk outside 240$''$. The inner disk contains $3.3
\cdot 10^{9}\ M_{\odot}$ of \hi, or 72\% of the total \hi\ mass of the
system.  The global profile of the inner disk is compared with that of
the entire disk in Fig.\ \ref{fig:glob}. Although not entirely
symmetric, it is clear that in the inner disk the difference between
approaching and receding sides is smaller than for the entire
disk. All modeling described below uses the $30''$ tapered cube. The
emphasis in this paper is on the geometry of NGC 4414 and we will not
analyse the dark matter content of the system. The outer disk is too
asymmetric for such an analysis; for the inner disk we refer to
\citet{vallejo03}.

\subsection{The inner disk}

\subsubsection{Tilted-ring fits}

\begin{figure*}
\includegraphics[width=0.9\hsize]{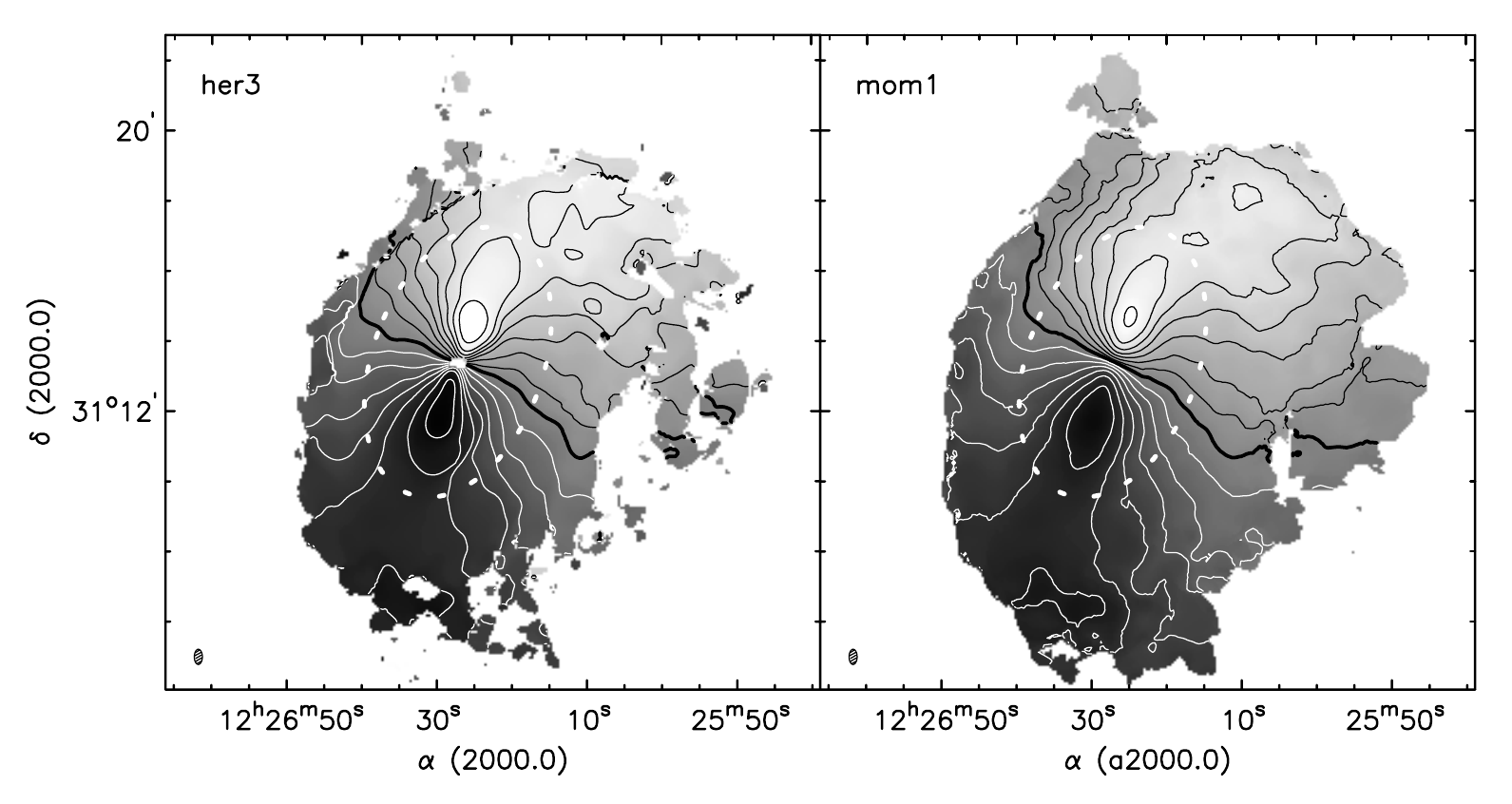}
\caption{Left: Hermite velocity field based on the $30''$-tapered
  cube. The thick black countour shows a systemic velocity of 711.5 km
  s$^{-1}$. Black contours then decrease in steps of 25 km s$^{-1}$,
  white contours increase in steps of 25 km s$^{-1}$. The white dashed
  ellipse with a major axis radius of $240''$ delimits the inner and
  outer disk. Right: intensity-weighted first-moment map. Contours as
  in left panel.
\label{fig:hermite}}
\end{figure*}

In addition to the moment maps we also created a Hermite velocity
field from the $30''$ tapered cube. We fitted third-order Hermite
polynomials to the \hi\ velocity profiles, retaining only those
profiles where the peak flux was higher than $3\sigma$, the velocity
dispersion larger than one channel width, and the velocity of the peak
of the profile within the velocity range of the galaxy. The integrated
\hi\ map was additionally used as a mask to remove spurious fits
outside the galaxy disk. The Hermite velocity field is shown in
Fig.\ \ref{fig:hermite} alongside the first-moment map.  Inspection of
the velocity fields shows the different extent of the Hermite field
compared to the moment map. This is due to the rejection criteria used
for the construction of the Hermite velocity field. The velocity
contours are also different, indicating the presence of non-Gaussian
profiles in a significant fraction of the disk. This, of course, is
the reason we are using the Hermite field: in the presence of
asymmetric profiles, it gives a better description of the rotation of
the bulk of the gas than an intensity-weighted first-moment map. See
\citet{db08} for an extensive discussion on this.

We first derive a rotation curve in the conventional manner using the
Hermite velocity field. We use the GIPSY task ROTCUR for a tilted-ring
fit. This task assumes that the gas rotates as a set of concentric
rings, each with their own inclination, position angle, and rotation
velocity.  Parameters of these rings are varied until the model
velocity field shows a sufficiently good match with the observed
one.  Usually the number of free parameters in tilted-ring
models is too large for all parameters to be determined
simultaneously. In practice, sets of parameters are progressively
fixed, and models with an increasingly smaller number of free
parameters are used to iterate to the optimum fit. See \citet{db08}
and \citet{gentile3198} for an extensive description of this
procedure.

\begin{figure*}
\includegraphics[width=0.9\hsize]{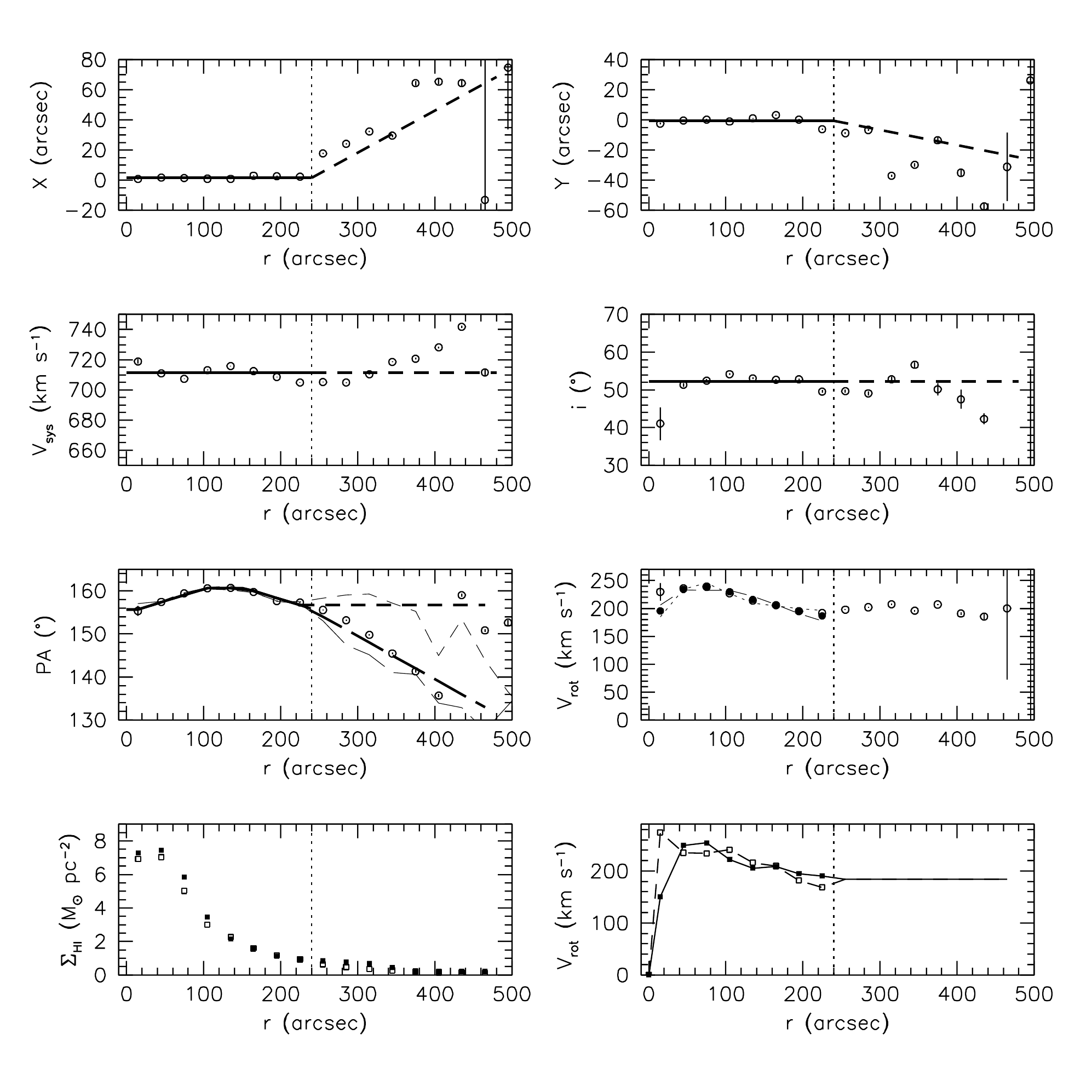}
\caption{Summary of NGC 4414 models. In all panels the vertical dotted
  line delimits the inner and outer disk.  {\bf Top three rows, inner
    disk:} open circles indicate the tilted ring parameters derived
  from a ROTCUR fit with all fit parameters free as described in
  Sect.\ 3.1.1.  The thick, full lines indicate the final model
  chosen for the inner disk.  In the third row, right-hand panel, the
  filled circles indicate the resulting final ROTCUR model for the
  rotation curve of the inner disk. The long-dashed and dotted curves
  indicate the receding and approaching side rotation curves of the
  inner disk, respectively. {\bf Top three rows, outer disk:} The
  dashed thick lines in the outer disk part of the plots indicates the
  values chosen for the final TiRiFiC model. In the left-hand panel of
  the third row the thick long-dashed line indicates the PA
  distribution for the approaching side, the short-dashed thick line
  that of the receding side. The thin long- and short-dashed lines
  show the PA values of the two sides when a ROTCUR fit is made to the
  velocity field with only PA and $V_{\rm rot}$ as free parameters.
  {\bf Bottom row:} the left-hand panel shows the \hi\ radial surface
  density distribution. Open squares indicate the approaching side,
  filled squares the receding side. The right-hand panel in the bottom
  row shows the approaching and receding sides rotation curves as open
  and filled squares, respectively. The horizontal line in the outer
  disk indicates the flat rotation curve assumed in the TiRiFiC
  modeling.
  \label{fig:innercurve}}
\end{figure*}

We first run ROTCUR with all parameters free (position of center $X,
Y$, systemic velocity $V_{\rm sys}$, position angle PA, inclination
$i$, and rotation velocity $V_{\rm rot}$). At this stage we assume the
rotation is purely circular, with no radial motions. We choose rings
with a width of $30''$ with the largest ring having a radius of
$500''$.  The resulting values of the tilted ring parameters as a
function of radius are shown in Fig.\ \ref{fig:innercurve} (top three
rows). It is immediately obvious from the Figure that most parameters
are well-behaved up until a radius between $\sim 250''$ and $\sim
300''$, before starting to show strong trends or large scatter at
larger radii.  The position of the center is most restrictive in that
regard, both the $X$ and $Y$ coordinates start to deviate from a
constant value around $R \sim 250''$.  NGC 4414 does not show a
compact central continuum source that would further constrain the
position of the center.

The rest of this sub-section deals exclusively with the inner disk.  Based
on the first model, we determine the mean $X$ and $Y$ position of the
center of the inner disk. We find a position $(X,Y)$ of the dynamical
center corresponding to $(\alpha, \delta) (2000.0) = (12^h26^m27.07^s,
31^{\circ}13'23.3'')$. The scatter in these parameters is $0.98''$ and
$2.76''$ in $X$ and $Y$, respectively. The position of the dynamical
center of the inner disk is therefore unambiguous and
well-determined. This is illustrated in the top panels of
Fig.\ \ref{fig:innercurve}.  This position corresponds closely to the
central position of the bulge as determined from archival HST WFPC-2
imaging, namely $(\alpha, \delta) (2000.0) = (12^h26^m27.19^s,
31^{\circ}13'22.9'')$, a difference of only $0.12^s$ or $1.5''$ in
right ascension, and $0.4''$ in declination.

Fixing the position of the dynamical center and an additional ROTCUR
run gives a value for the systemic velocity $V_{\rm sys} = 711.5 \pm
4.6$ km s$^{-1}$ (second row, left panel in
Fig.\ \ref{fig:innercurve}). Fixing $V_{\rm sys}$ in turn, we find a
constant inclination of $i=52.3^{\circ} \pm 1.5^{\circ}$ (excluding
the innermost points; second row, right panel in
Fig.\ \ref{fig:innercurve}).  We checked for differences between
values for the approaching and receding side, but find almost
identical results: $i=52.7 \pm 1.3$ for the receding side, and
$i=52.4\pm 2.5$ for the approaching side. So far, the inner disk of
NGC 4414 thus seems to be a very well-behaved, regularly rotating disk
without any major asymmetries.

The values for PA show some radial variation: the innermost ring gives
PA $\simeq 155^{\circ}$, which then increases to $\sim 161^{\circ}$ at
$R\sim 125''$, and decreases again to $\sim 155^{\circ}$ at $R=240''$
(third row, left panel in Fig.\ \ref{fig:innercurve}). In light of the
large apparent differences in the velocity field of the outer disk, we
explore whether the trend in PA is the result of differences between
the approaching and receding sides of the inner disk. We made ROTCUR
models of the approaching and receding sides separately, using the
parameters given above, and with PA and $V_{\rm rot}$ as free
parameters. The radial variation of PA for both sides is remarkably
similar in the inner disk: we find almost identical trends, with a
mean absolute difference between the PA values of only $\sim
0.6^{\circ}$, with the largest difference being $1.4^{\circ}$ for the
innermost ring (also shown in the third row, left panel in
Fig.\ \ref{fig:innercurve}). 

We describe the PA behavior in the inner disk with a simple
three-line-piece model, and with all parameters except $V_{\rm rot}$
fixed, we derive the final tilted-ring rotation curve of the inner
disk. This is shown in the right-hand panel in the third row in
Fig.\ \ref{fig:innercurve}. There is a the striking resemblance with the
curve derived with all parameters free. The only significant
difference is in the innermost point, where the unconstrained
inclination for the all-free curve resulted in a higher rotation
velocity. The outer rotation curve remains flat. We will
return to this when discussing the outer disk.

\begin{figure*}
\includegraphics[width=0.9\hsize]{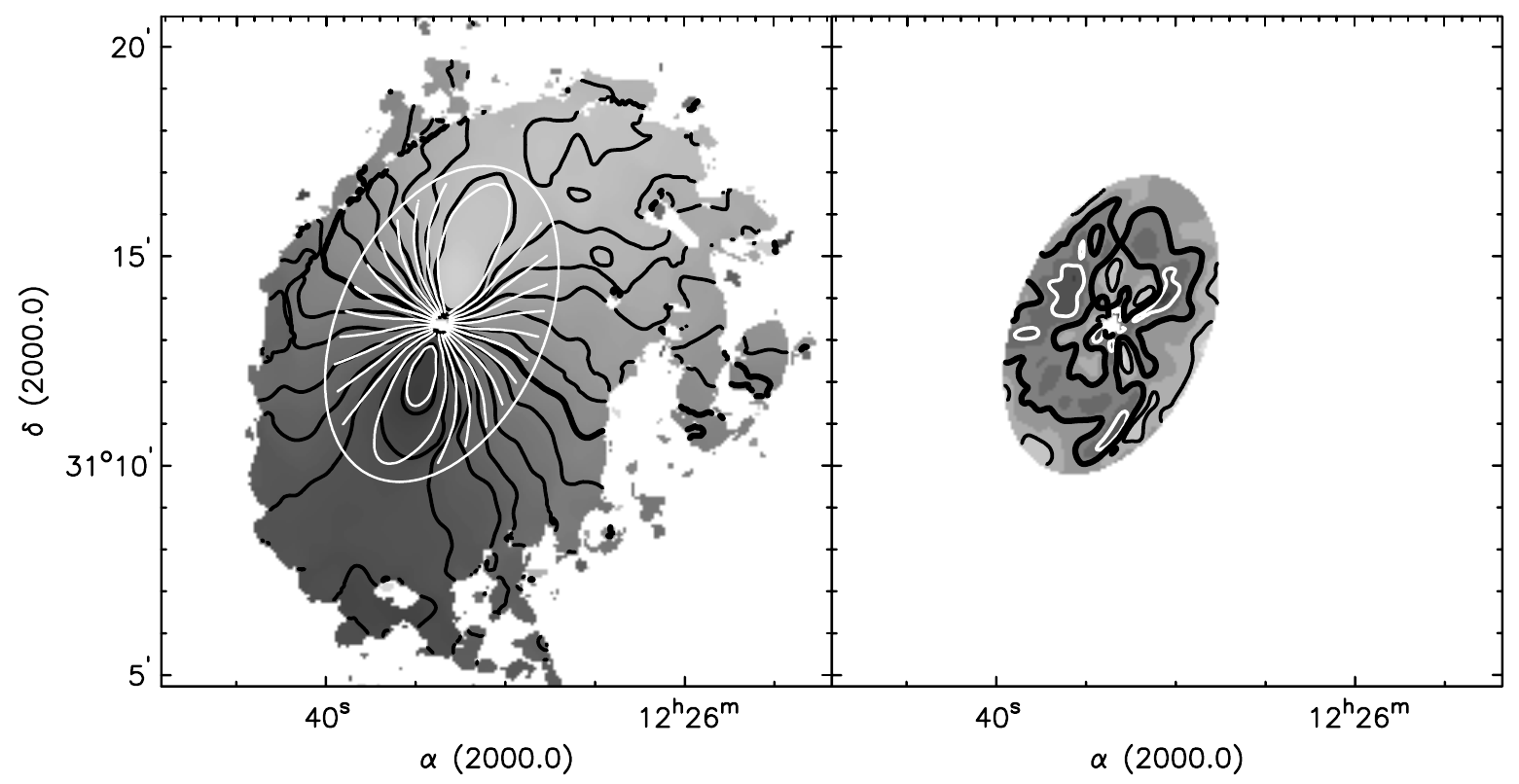}
\caption{Left panel: the Hermite velocity field (grayscale with black
  contours) with the best-fitting model velocity field for the inner
  disk overlaid. Contours are as in Fig.\ \ref{fig:hermite} with a
  contour spacing of 25 km s$^{-1}$. Right panel: residual velocity
  field derived by subtracting the model from the observations. The
  thick black contour shows the 0 km s$^{-1}$ level. The thin black
  contour $+10$ km s$^{-1}$, the thin white contour $-10$ km
  s$^{-1}$.The maximum and minimum values found are $+16$ and $-16$ km
  s$^{-1}$, respectively.\label{fig:innervelfi}}
\end{figure*}

Using the final rotation curve we build a model velocity field, and
compare this with the observed one in Fig.\ \ref{fig:innervelfi}. The
overlay on the model velocity field on the Hermite velocity field
shows good agreement. Small local deviations exist, but the global
properties of the velocity field of the inner disk are captured
well. We also show the residual velocity field (derived by subtracting
the model from the observations), and find no large-scale systematic
deviations there.  The average value of the residual velocity field is
$-0.4$ km s$^{-1}$ with an rms spread of $6.2$ km s$^{-1}$. The median
of the absolute value of the residuals is $4.4$ km s$^{-1}$ or just
over one channel spacing.

The rotation curves of the approaching and receding sides separately
are also shown in Fig.\ \ref{fig:innercurve} (third row, right-hand
panel). They show the same global behavior, namely the sharp rise in
the center followed by the gentle decline out to $R\sim 240''$. The
mean difference between both curves (defined as approaching minus receding)
is $+0.1 \pm 13.4$ km s$^{-1}$ measured over the inner disk. The mean
absolute difference is $10.5 \pm 7.3$ km s$^{-1}$. The lack of global
residual patterns in Fig.\ \ref{fig:innervelfi}, and the good
agreement of the model velocity field with the data, suggests that the
differences are mainly caused by local velocity deviations, and not by
global asymmetries.

\subsubsection{TiRiFiC fit}

The inner disk of NGC 4414 is well-behaved without major asymmetries
and disturbances, and, as shown, can be very well described using a
tilted-ring model. We now use this tilted ring model as input for more
sophisticated modeling using the Tilted Ring Fitting Code (TiRiFiC)
\citep{tir07}.  TiRiFiC distributes cloud particles through a
user-created three-dimensional (right ascension, declination,
velocity) data cube, using user-defined geometric and kinematic
parameters as distribution functions. Among the parameters that can be
defined are the rotation curve, radially varying inclination and
position angles, the surface density profile, the vertical
distribution and extent of the gas, and the velocity dispersion. By
convolving with the beam of the observation, TiRiFiC produces a data
cube that can be ``observed'' and treated like the original
observation. 

We again limit the modeling to the inner disk and model the
approaching and receding sides of the disk separately. We use the
tilted ring orientation parameters derived above to derive
\hi\ surface density profiles of the approaching and receding sides
using the ELLINT task in GIPSY. These profiles are used as
additional input for TiRiFiC and are shown in
Fig.\ \ref{fig:innercurve}.

We model the galaxy with a single thin disk component. A number of
exploratory runs of TiRiFiC confirms the results from the ROTCUR runs:
the inclination and position angle values are well-behaved, and for
all practical purposes the values of these parameters are equal to the
ones derived previously. In subsequent runs we therefore fixed $i$ and
PA to the ROTCUR values.

The gas velocity dispersion was determined using TiRiFiC and we find a
value of 11 km s$^{-1}$. This number is not corrected for instrumental
velocity resolution (channel spacing), which would bring the number
down to 10 km s$^{-1}$. We assume a vertical sech$^2$ profile and find
a vertical scale height of $z_0 = 0.25$ kpc, although this value is
uncertain. Scale heights between $\sim 0.1$ and $\sim 0.5$ kpc give
similar results.

The \hi\ surface density profile, which was kept fixed in the fits so
far, was then allowed to vary (keeping all other parameters fixed). We
tested models where the surface density profile was left completely
free, where the profiles were allowed to vary using one single scale
factor for the entire profile, and one where corresponding radii from
approaching and receding side were allowed to vary in tandem. In the
end we found the best solutions for the model with the constant scale
factor, with surface densities differing only a few percent from their
input values. These are the models presented in the rest of the paper.

\begin{figure*}
\includegraphics[width=0.9\hsize]{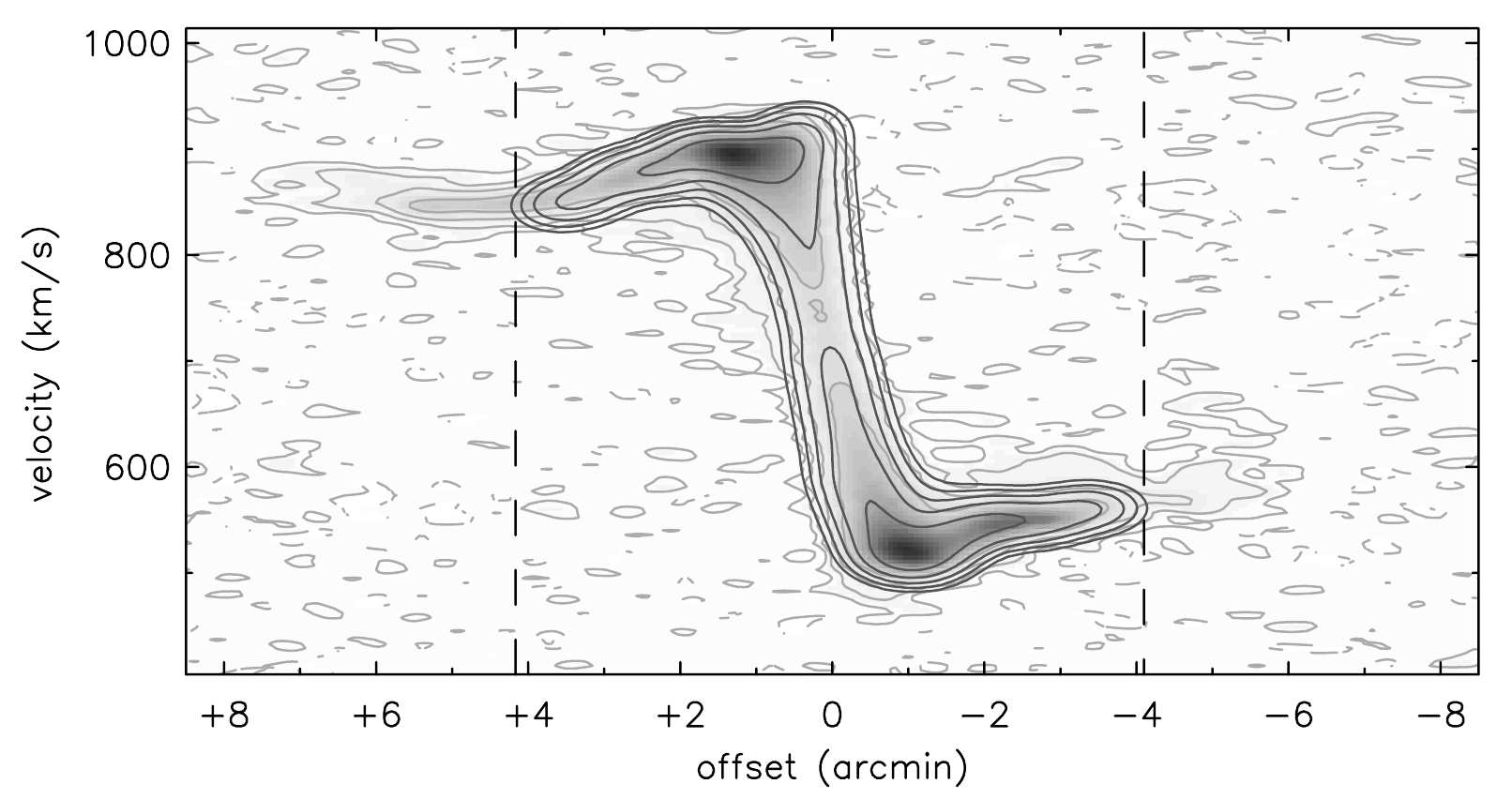}
\caption{Overlay of the best TiRiFiC model of the inner disk on the
  major axis pV diagram. The data is shown in gray-scale and
  light-gray contours. The model in black contours. For both, the
  lowest contour level shown is $1.5\sigma$. Subsequent contour levels
  each increase by a factor 2.5.  For the data, the dashed contours
  also show the $-1.5\sigma$ level.  The vertical dashed lines delimit
  the inner disk. \label{fig:majaxis}}
\end{figure*}

\begin{figure*}
\includegraphics[width=0.9\hsize]{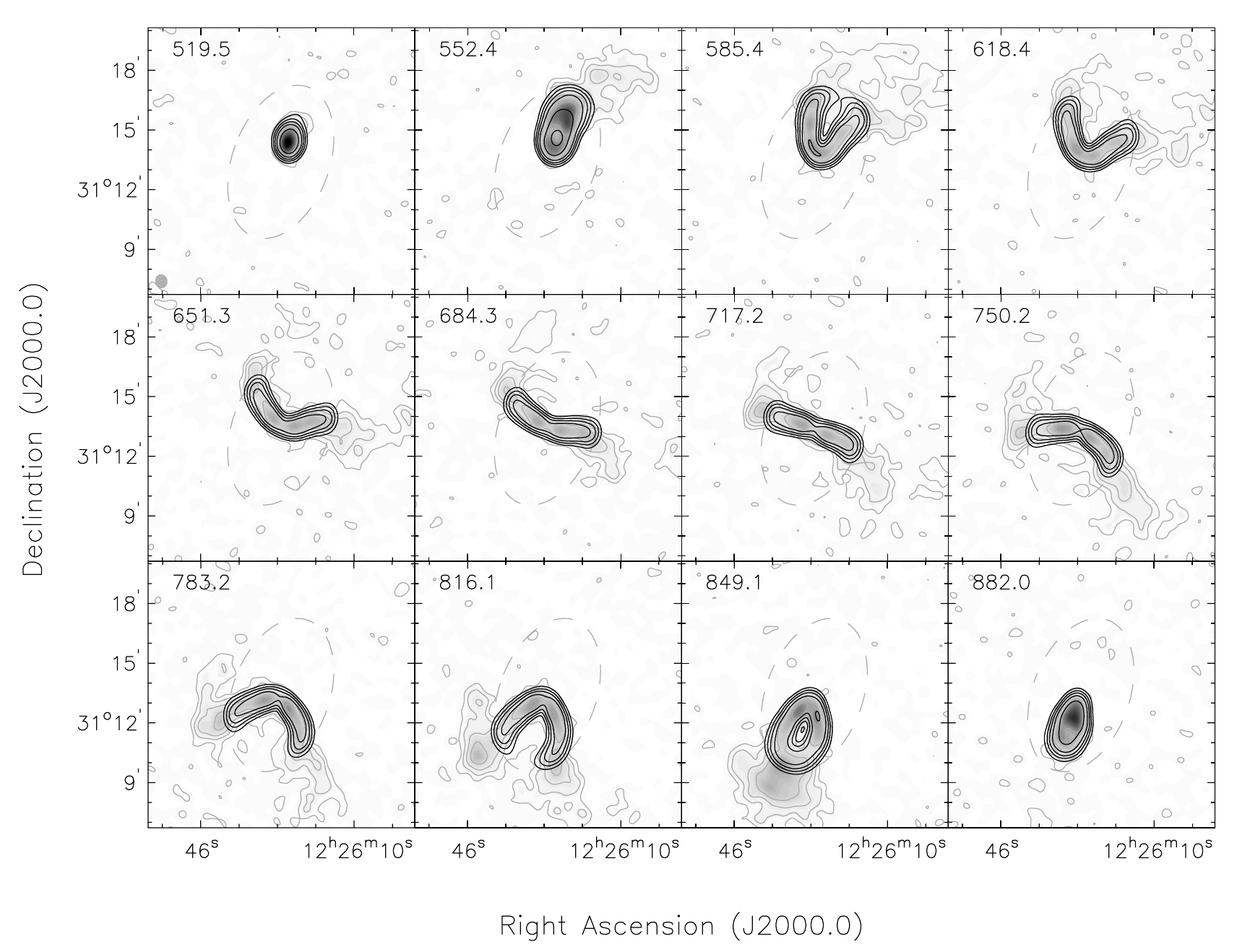}
\caption{Overlay of the best model of the inner disk on selected
  channel maps. Contour levels for both data and model are $(2, 5, 10,
  20, 50) \cdot \sigma$ where $\sigma=0.23$ mJy beam$^{-1}$ or $0.8
  \cdot 10^{18}$ cm$^{-2}$. The dashed ellipse denotes the inner disk.
\label{fig:chans_inner}}
\end{figure*}

The parameters of this best-fitting single disk model are shown in
Fig.\ \ref{fig:innercurve}. An overlay of the model on the major axis
position-velocity diagram is shown in Fig.\ \ref{fig:majaxis}, while
in Fig.\ \ref{fig:chans_inner} we compare selected model channel maps
directly with the data.

The distribution and kinematics of the \hi\ in the inner disk of NGC
4414 are well described by this simple single-disk model. We only see
evidence of small amounts of ``beard'' gas (i.e., gas at velocities
lower than the local rotation velocity) that is not captured by this model.
Compare for example our observations with those of
NGC 3198, presented in \citet{gentile3198} (their Fig.\ 4), where 
the ``beard'' is much more prominent.

In Fig.\ \ref{fig:majaxis} one can see that on the receding side the
``beard'' component is present up to a radius of $\sim 200''$. On the
approaching side, we see a small amount of gas associated with
the rising part of the rotation curve (``beard''), as well as an
additional component between $-2.5'$ and $-4'$.  We interpret these
components as extra-planar gas and determine its mass in several ways.

First, we determine the total flux in the best-fitting model for the
inner disk, and compare this with the observed inner disk flux. The
model results in an \hi\ mass of $2.9 \cdot 10^{9}\ M_{\odot}$,
compared with an observed inner disk \hi\ mass of $3.3 \cdot
10^{9}\ M_{\odot}$. The difference is $4 \cdot 10^{8}\ M_{\odot}$, or
12 per cent of the inner disk mass. A possible disadvantage of this
method is that the radially averaged \hi\ surface density has been
used as an input for the model, and it is not clear how this affects
the comparison with the non-averaged observed surface densities.  This
value is therefore best regarded as an upper limit.

A second estimate can be derived by noting that in
Fig.~\ref{fig:majaxis} the extra-planar gas has a predominantly low
column density, while the part of the galaxy described by the model
mostly has higher column-density values.  We can therefore blank all
high-column density values in the data cube, and measure the mass of
the remaining low-column density gas by blanking in the data cube all
pixels in the cylinder describing the inner disk that are brighter
than 0.46 mJy beam$^{-1}$ (the lowest contour of the model plotted,
corresponding to a 2$\sigma$ level in the data). We find an \hi\ mass
of $2.1 \cdot 10^{8}\ M_{\odot}$, or 6.5 per cent of the total
\hi\ mass of the inner disk.  It is possible that this is a slight
overestimate, it may still contain emission from faint wings from
bright inner disk profiles.

A third way to derive the amount of extra-planar gas is to use the
method described in \citet{fraternali2002} where Gaussian fits are
made to velocity profiles in a data cube where in each profile, values
less than 30 per cent of the respective peak value have been
blanked. These Gaussian fits thus describe the bright cores of the
profiles.  Subtracting these from the original, non-blanked profiles,
results in a data cube containing only the emission from broad,
low-level wings, secondary components etc.  Using this method on the
entire disk, we find that prominent extra-planar gas components are
only present in the inner disk. The method does produce some low-level
fitting artefacts, so we use several different flux cut-offs to gauge
the reliability of the emission. We find the extra-planar gas in the
inner disk to be $\sim 2.7$ to $\sim 4.0$ per cent of the mass of the
inner disk, or $\sim 2.0$ to $\sim 3.0$ per cent of the total
disk. These numbers should probably be regarded as a lower limit, as
gas which is extra-planar with no corresponding main-disk gas at the
same position (such as gas in a warp) will not be accounted for by
this method.

Although not the topic of this subsection, this method also allows an
estimate of the amount of ``extra'' gas in the outer disk. We call it
``extra'', as it is not clear whether this outer disk gas is strictly
speaking ``extra-planar'' due to the disturbed nature of the outer
disk. We find that $\sim 1.0$ to $\sim 2.0$ per cent of the total
\hi\ mass is in the form of ``extra'' outer disk gas.  This gas, in
contrast with the inner disk, is distributed at a low level throughout
the outer disk, and it is very well possible that this happens at a
level below the accuracy allowed by this method.

We explored a fourth method, which uses the Hermite velocity field to
shift all \hi\ profiles in velocity so their peaks all align at a
common reference velocity. Blanking the velocity channels closest to
this reference velocity automatically isolates the high-velocity
emission, and would thus indicate the importance of the extra-planar
gas.  In practice, however, beam smearing broadens the innermost
profiles, making the method unusable.

In summary, we find that a single, regularly rotating disk model is a
good description of the inner disk of NGC 4414. Depending on the
method used we find that 12 (uncertain upper limit) or 6.5 (upper
limit) or 2--3 percent (lower limit) of the total \hi\ mass in the
inner disk is associated with an extra-planar component.  An estimate
of around 4 per cent thus seems appropriate. Given the high
star-formation rate and location coinciding with the optical disk, it
is likely that most, if not all, of the extraplanar gas above the
inner disk is associated with star formation processes.

\subsection{The outer disk}

While the inner disk has a regular morphology and is symmetrical, this
is clearly not the case for the outer disk. The \hi\  disk is more
extended towards the south and to the northwest. In the northern
part of the disk we also find what look like arms or streams. In the
eastern part of the disk the column density contours are closer
together, suggesting either a sharp edge to the \hi\  distribution, or
one that is bending into the line of sight.

The Hermite velocity field shows a similar wealth of structures.  In
the outer disk the PA values of the approaching and receding sides
diverge sharply: the PA of the receding side remains more or less
constant at $\sim 155^{\circ}$ (but with a large scatter), while that
of the approaching side drops sharply to values of $\sim 130^{\circ}$
at the outermost radii.  This behavior clearly justifies our division
in an inner and outer disk.  Kinks in the velocity contours towards
the edge of the disk indicate either abrupt changes in the geometrical
parameters, or the presence of non-circular motions.

Many of these features can be identified in the channel maps as well
(Fig.~\ref{fig:chans}). The strong position angle twist is visible
between $V \sim 550$ and $\sim 600$ km s$^{-1}$. The kinks in the
velocity contours in the eastern part of the disk show up as
low-column density components perpendicular to the main \hi\ structure
in the channel maps between $V \sim 650$ and $\sim 800$ km s$^{-1}$.

The asymmetry in the disk and the presence of localized structures
already indicate that it will be more difficult to capture these
details in a simple, symmetrical tilted-ring model, than was the case
with the inner disk, especially as our goal here is to describe the
system without adding a large number of \emph{ad hoc} local
components.  We will therefore construct a model that reproduces the
large-scale features of the outer disk, but we will not attempt to try
and incorporate every small-scale feature in this description. The
disturbed nature of the outer disk also means an unambiguous
separation between planar and extra-planar gas (even if such a
distinction is meaningful for this particular case) will be
difficult, if not impossible, to achieve, and we will not attempt to
do so here.  All modeling is again done with the TiRiFiC program.

As the geometry of the outer disk is much less constrained, we first
make a number of assumptions. We assume that the outer disk has a flat
rotation curve with a rotation velocity of 184 km s$^{-1}$, a constant
systemic velocity of 711.5 km s$^{-1}$, and a constant inclination of
52.3$^{\circ}$. These last two values are also the
values adopted for the inner disk. These assumptions are not
unreasonable as Fig.\ \ref{fig:innercurve} shows.  We assume the \hi\ 
surface density profiles also shown in Fig.\ \ref{fig:innercurve}
(bottom-left panel). We model the approaching and receding sides
separately. For the inner disk we retain the model derived in the
previous section.  We define the outer disk as the part of the galaxy
between $240'' < R < 480''$. This covers most of the disk, except for
a small number of isolated \hi\  features at larger radii which we do not
attempt to capture in the model. The small filling factors of these
potential outer rings would make any model be of limited value.

The models presented here were not fitted to the data in a strict
sense. The large number of free parameters and asymmetric nature of
the disk make this a challenge. Models were therefore mostly
adjusted and evaluated by eye, by careful comparisons of overlays on
channel maps and position velocity slices along various axes. Here we
do not present all models investigated, but  only the ones that lead
to an acceptable description of NGC 4414.

As a compact way of illustrating features of the different models, we
plot selected channel maps where we show the models overlaid on the
data.  As stated earlier, the actual evaluation of the
models was done using a much larger set of overlays and slices.

The first model (\emph{`pa'}) incorporates the changes of PA with
radius mentioned earlier.  Approaching and receding sides show a
distinctly different behavior: for the receding side we keep the
value constant at PA $=156.7^{\circ}$, which is the outermost value
found in the inner disk. The approaching side shows a strong PA twist
and here we change the PA linearly from $156.7^{\circ}$ at the
outermost radius of the inner disk to $133^{\circ}$ at the outer
radius of the outer disk. This is illustrated in
Fig.\ \ref{fig:innercurve}.

%\begin{figure*}
%\plotone{renzogram.pdf}
%\caption{Renzograms of the data (left panel) and models (three
%  rightmost panels). Light-gray contours show the emission at 581.3 km
%  s$^{-1}$ (top) and 845.0 km s$^{-1}$ (bottom). Black contours show
%  emission at 680.2 km s$^{-1}$ (top) and 779.0 km s$^{-1}$
%  bottom. For each velocity shown the contour levels are $(3, 10,
%  20)\cdot \sigma.$ Models are described in the
%  text.\label{fig:renzo}}
%\end{figure*}

\begin{figure*}
\includegraphics[width=0.9\hsize]{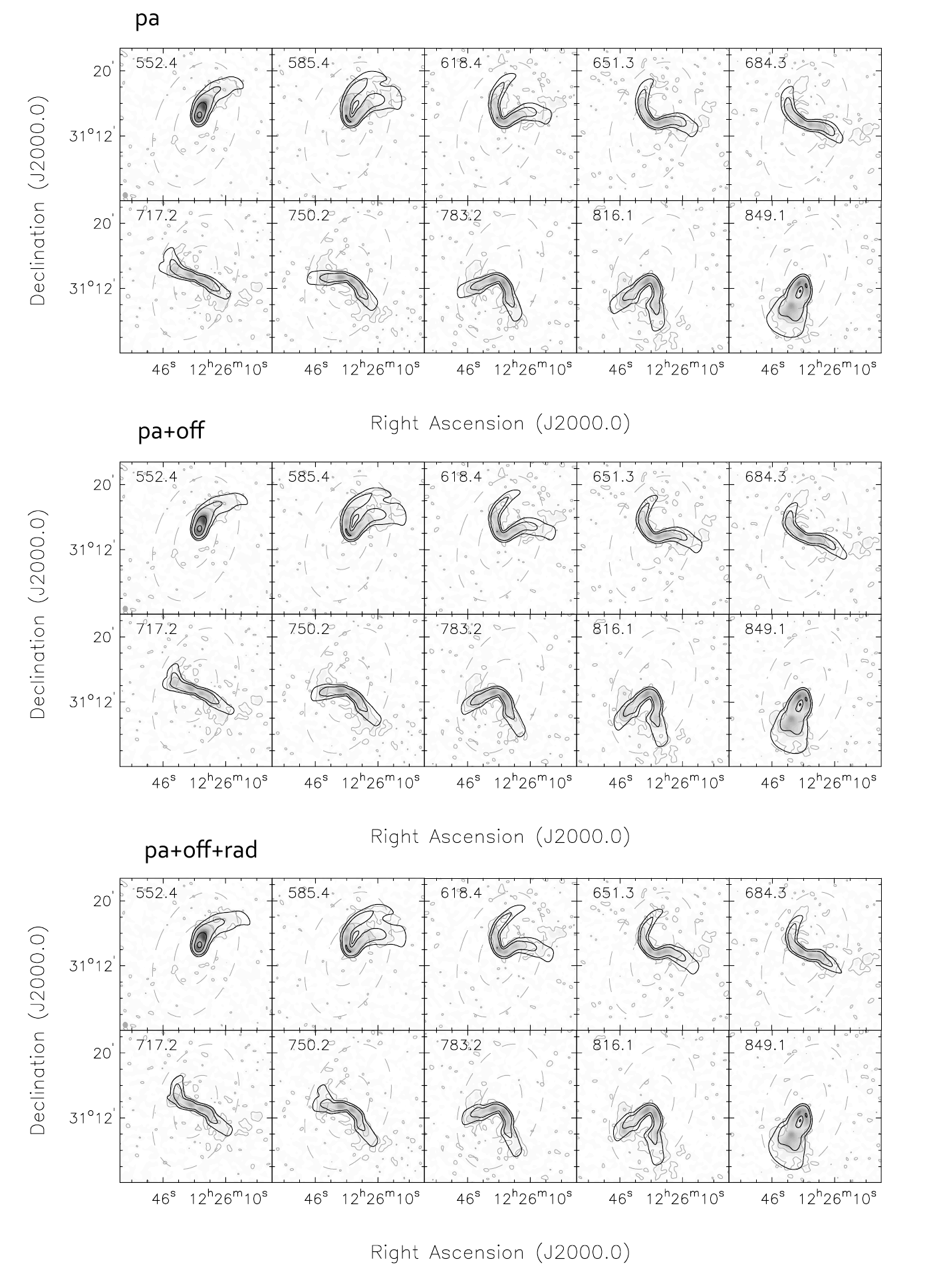}
\caption{Overlay of the models on selected channel maps. Contour
  levels for both data and model are $(2, 10, 50) \cdot \sigma$ where
  $\sigma=0.23$ mJy beam$^{-1}$ or $0.8 \cdot 10^{18}$ cm$^{-2}$. The
  two dashed ellipses in each panel denotes the outer radii of the
  inner and outer disk. Details of the models are described in the text.
\label{fig:outermodels}}
\end{figure*}

In the top panel of Fig.\ \ref{fig:outermodels} we compare the data
and model `\emph{pa}'. The model still has a number of flaws. The
emission in the eastern part of the disk extends too far radially and
does not have the sharp drop-off observed in the data.  This is
clearly visible in the channel maps at $V=717.2$ and $750.2$ km
s$^{-1}$. The model emission in the southern part of the disk is also
offset azimuthally clock-wise from the observed emission (visible at
at $V=816.1$ and $849.1$ km s$^{-1}$). A similar offset is seen in the
northern outer part at $V=585.4$ km s$^{-1}$, but here the model
emission is offset azimuthally counter-clockwise from the observed
emission.

We found that most of these flaws can be rectified by a
systematic change in the position of the dynamical center with
radius. We incorporate this in the next model (`\emph{pa+off}') and
change the position of the dynamical center of both sides of the disk
linearly as a function of radius from $(\alpha, \delta) (2000.0) =
(12^h26^m27.07^s, 31^{\circ}13'23.3'')$ at $R=240''$ to $(\alpha,
\delta) (2000.0) = (12^h26^m24.46^s, 31^{\circ}13'58.8'')$ at
$R=480''$. This corresponds to a change of $-67''$ in right ascension
and $-24''$ in declination, or a maximum shift of $71''$ towards a
position angle of 236$^{\circ}$. At the distance of NGC 4414 this
shift corresponds to 6.1 kpc (as projected on the sky; further
interpretation of this shift is given in Sect.\ 4). Additional
modeling shows that the magnitude of the shift is constrained to about
$\pm 10''$.  The effect of this shift on the model is a compression of
the column density contours towards the east and an extension of the
\hi\ distribution towards the west, in better agreement with the
data. The model and data are compared in the middle panel of
Fig.\ \ref{fig:outermodels}. The emission now extends farther west, is
more compressed towards the east, and the azimuthal offsets in the
north and south have mostly disappeared.

The model does not yet adequately describe the ``trailing'' and
``leading'' \hi\ emission seen at the outer edge of the disk in the
channel maps around $V=750$ km s$^{-1}$ (eastern edge) and $V=700$ km
s$^{-1}$ (western edge) in Fig.\ \ref{fig:chans}.  This turns out to
be more difficult to model with standard tilted ring parameters, and
necessitates the introduction of radial velocities.  We introduce a
radial velocity term of $-20$ km s$^{-1}$ for all rings in the outer
disk (model `\emph{pa+off+rad}').  The model is also shown in
Fig.\ \ref{fig:outermodels}.  The difference is most clearly seen at
$V=750.2$ km s$^{-1}$. The PA of the outer emission at $V=717.2$ km
s$^{-1}$ is also better described, as is the slight lengthening of the
northern emission ``arm'' at $V=651.3$ and $V=684.3$ km s$^{-1}$ and
the slight PA change in the western ``arm'' in the same channels.  If
we assume that the spiral arms in NGC 4414 are trailing arms, then the
eastern side of the disk is closest to us. In that case a negative
radial velocity as modeled here corresponds to an inflow.

As noted before, due to the asymmetries in the disk, not all
small-scale features can be captured in a tilted-ring model, nor do we
model the outermost western features.  The two most important features
not modeled are the \hi\ clumps visible on the western side of the
disk in the channel maps between 618.4 and 717.2 km s$^{-1}$.  The
integrated \hi\ map in Fig.~\ref{fig:moms} shows that these form an arm
or tail that is present at large radius over a limited range in
position angle.  The second feature not modeled is present between
750.2 and 816.1 km s$^{-1}$ on the eastern side. This is caused by the
difference in PA distribution between the approaching and receding
side. The model for the receding side has a constant PA in the outer
disk (in order to model the outer disk around 850 km s$^{-1}$), while
the approaching side has a twist in the PA (in order to model the
outer \hi\ at 585.4 km s$^{-1}$). The same twist does also capture the
change in PA visible in the outer disk at $717.2$ and $750.2$ km
s$^{-1}$ in Fig.\ \ref{fig:outermodels}. We would be able to fit the
similar features visible in the channels at $V=783.2$ and $V=816.1$ km
s$^{-1}$ in Fig.\ \ref{fig:outermodels} by applying the PA twist
there. The reason for not doing this is that it would worsen the model
fit on the western side if the disk, where no such strong PA
variations are seen. 

This model still assumes constant inclination. We ran a number of
models to check whether this assumption still held, or whether
non-constant values could improve the result. With inclination as a
free parameter, the average value of $i$ for the approaching side
changes by $\sim 1^{\circ}$, with a large scatter. The receding side
prefers a somewhat lower value of $\sim 45^{\circ}$, although this does
not result in an increase in quality of the models, improving the fit
in one channel map, while worsening it in others. No gains are
therefore made by changing the inclination, and we keep it fixed at a
constant value.

Finally, in the next Section, we will interpret our best model as
showing the presence of a U-shaped warp. In such a model the systemic
velocity could potentially vary with radius. We have tested a number
of models where the outer disk has a different systemic velocity than
the inner disk. We find that models where the outer systemic velocity
is higher than the inner systemic velocity do not perform well. A
lower outer systemic velocity is however possible, down to a minimum
value of $\sim 695$ km s$^{-1}$. The discriminatory power of the
models is not great, however, and we proceed to keep using the
constant $V_{\rm sys}$ model.

In summary, the outer disk can be reasonably modeled by introducing a
small shift in the position of the dynamical center, and the addition
of small radial motions. All this indicates that the outer disk is
probably not in an equilibrium state, and confirms the conclusion
derived from the \hi\ morphology that NGC 4414 must have undergone or
is undergoing some kind of disturbance.

%\begin{deluxetable}{llcrccrl}
%\tabletypesize{\scriptsize}
% \tablecaption{Properties of template rotation curves.\label{tab:template}}
% \tablewidth{0pt}
%\tablehead{
%\colhead{$t$} & \colhead{$M_I$}  & \colhead{$\Delta M_I$} & \colhead{$V_0$} & \colhead{$r_{PE}/h$} & \colhead{$\alpha$} & \colhead{$h$} & \colhead{$V_{\rm max}$} \\
%\colhead{(1)} & \colhead{(2)} & \colhead{(3)} & \colhead{(4)} & \colhead{(5)} & \colhead{(6)} & \colhead{(7)} & \colhead{(8)} 
%} 
%\startdata
%0 &  $-23.8$ &  0.4          & 270   & 0.37    & 0.007    & 10.4  & 294 \\
%1 &  $-23.4$ &  0.4          & 248   & 0.40     & 0.006    & 8.6 & 267 \\
%9 &  $-19.0$ &  2.0          & 64   & 0.72     & 0.087    & 1.1  & 102 
%\enddata
%\end{deluxetable}

\section{Discussion}

\subsection{A U-shaped warp}

The previous section established that NGC 4414 can be characterized by
a symmetrical, very regular inner disk, and a more asymmetric,
disturbed outer disk, which, although mostly dominated by rotation,
shows evidence of radial motions and systematic shifts in the
position of the dynamical center.

This shift in dynamical center position occurs roughly along the minor
axis of the galaxy. This can be interpreted as a distortion of the
rings in the plane of the galaxy, but also as the projection on the
sky of an offset of the outer rings perpendicular to the plane of the
inner disk (i.e., in the $z$-direction). In other words, a U-shaped
warp.  This could also explain the sharp edge in the eastern part of
the disk, with the emission bending into the line of sight. The
increased second-moment values (due to the presence of multiple
velocity components) towards the NE part of the disk support this (see
Fig.\ \ref{fig:moms}).

Using the known inclination and a simple geometrical model, we can
translate the shift of the center $\Delta r$ on the sky into the
spatial offset of the rings perpendicular to the inner disk $\Delta r'
= \Delta r / \sin i$. The maximum shift in the plane of the sky (of
the outermost ring) is $\sim 6.1$ kpc. This translates into a maximum
offset perpendicular to the plane of $\sim 7.6$ kpc. This can be
compared with the diameter of the inner disk of 41.5 kpc, or the
diameter of the outermost ring of 83.0 kpc.  The latter value results
in a ratio of maximum vertical offset of the outer ring and the outer
ring diameter of $\sim 1:11$. The warp angle (angle between main disk
and a line connecting the center of the galaxy with the outermost
warped ring) is $\sim 10^{\circ}$. Comparing this with a study of
warps in edge-on disk galaxies by \citet{garciaruiz}, we find that NGC
4414 has a slightly higher than average warp angle, but that the value
is not exceptional.

%\begin{figure}
%\plotone{geometry.pdf}
%\caption{Oblique view of a ray-traced three-dimensional TiRiFiC
%  model. The viewing angle is $30^{\circ}$ horizontally with respect
%  to the line of sight.
%\label{fig:geom}}
%\end{figure}

One cause of U-shaped warps is thought to be the interaction with the
intergalactic (intra-cluster) medium through ram-pressure. 
Ram pressure effects are usually associated with denser environments,
such as clusters, and the question is whether the less dense
intra-group medium in Coma I can be responsible for these effects as
well. In a study of the \hi\ in NGC 300 in the Sculptor group,
\citet{westmeier} evaluate the conditions for ram pressure effects in
a group environment, and show that for a velocity with respect to the
intra-group medium of $> 200$ km s$^{-1}$ and a intra-group medium
density larger than a few times $10^{-5}$ cm$^{-3}$, the effects of
ram pressure should be visible in the outer parts of galaxies in these
group environments.

The interaction with the intra-group medium is modelled  in \citet{haan}
and used to explain the formation of warps. They model the movement of
a disk galaxy with an extended \hi\ disk through the intragroup medium
of a galaxy group with a mass of $\sim 10^{13}\, h^{-1}$ $M_{\odot}$,
the mass of the most commonly occuring galaxy groups (the
Coma I group has an estimated mass of $\sim 5 \cdot 10^{13}$
$M_{\odot}$; \citealt{kar11}).  \citet{haan} show that having a disk
galaxy move with respect to the intragroup medium at a velocity
corresponding to the three-dimensional velocity dispersion of the
group ($\sim 260$ km s$^{-1}$) and with a typical intragroup medium
density of $\sim 1\cdot 10^{-4}$ cm$^{-3}$, is enough to rapidly (within
two rotational periods of the galaxy) set up a warp-structure which is
long-lived (at least 10 rotational periods).  With these parameters
the gravitational force of the galaxy dominates the ram pressure
force, and a warp structure can exist without actual stripping of gas
occuring.

Without more accurate estimates for the peculiar velocity of NGC 4414
and the surrounding group medium it is, however, difficult to say anything more
quantitative about the ram pressure effects.

Lastly, the morphology of NGC 4414 could be caused by an interaction
with a neighboring galaxy. As NGC 4414 is part of a group environment
it is likely that (unless it is falling in the first time) it will
have undergone some interaction with other group members in the past.
The peculiar motions of galaxies in the Coma I region are however
complex \citep{kar11} and it is difficult to unambiguously define a
candidate galaxy that could have caused the current \hi\ morphology of
NGC 4414.

\subsection{Optical shells}

As described in Sect.\ 2.2, deep optical imaging has been obtained
from two separate observing campaigns, resulting in a $B$-band image
from KPNO and an $r'$-band image from the La Palma INT telescope.  The
independent observations allow us to gauge the reality of faint
features in the images, and in the following we only discuss features
present in both images.

\begin{figure*}
\includegraphics[width=0.9\hsize]{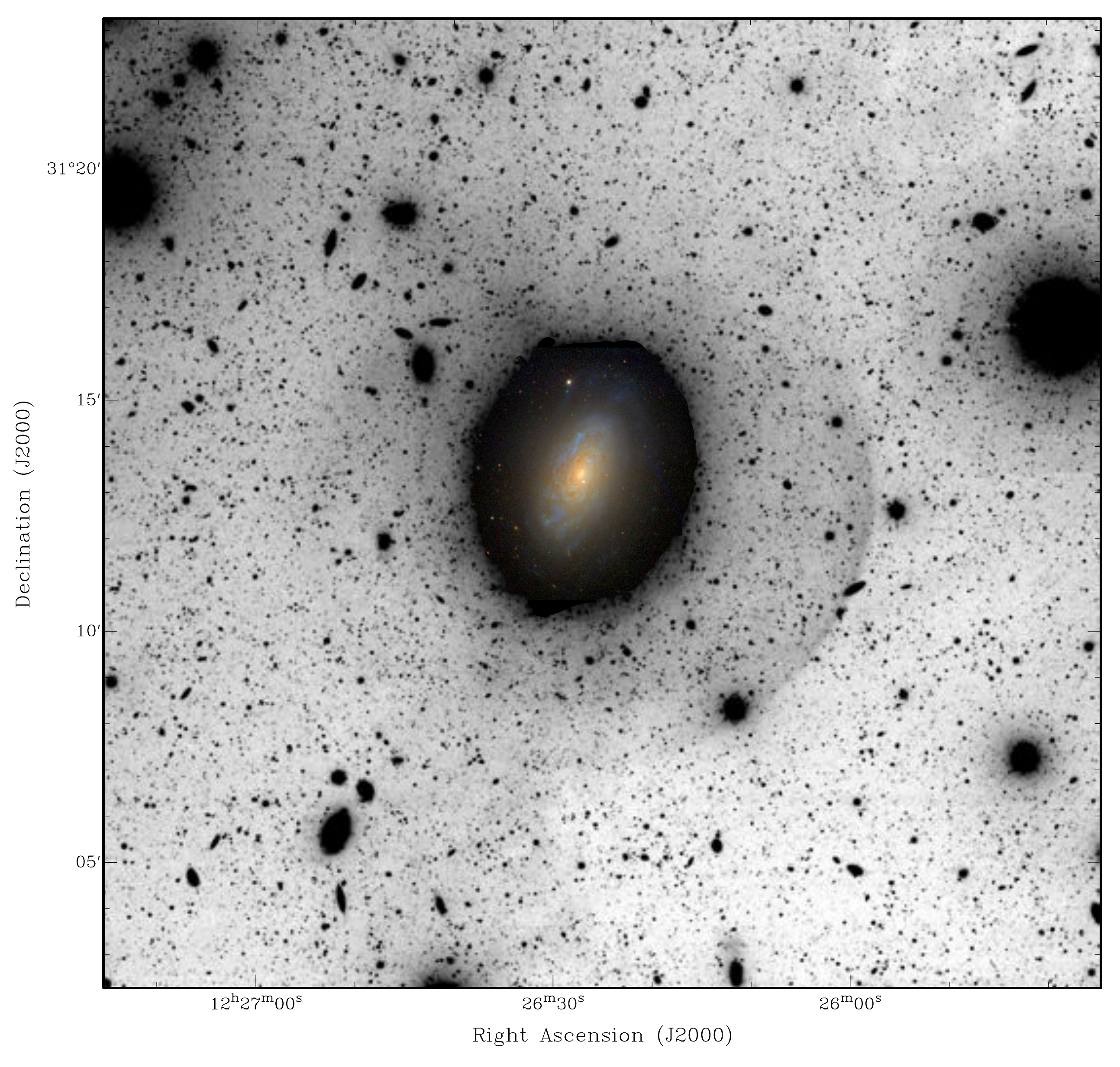}
\caption{Combination of the deep $B$-band image (grayscale)
  and a SDSS image (color). The $B$-band image is smoothed to $2''$
  resolution to enhance the shell features. \label{fig:sdss}}
\end{figure*}

\begin{figure*}
\includegraphics[width=0.9\hsize]{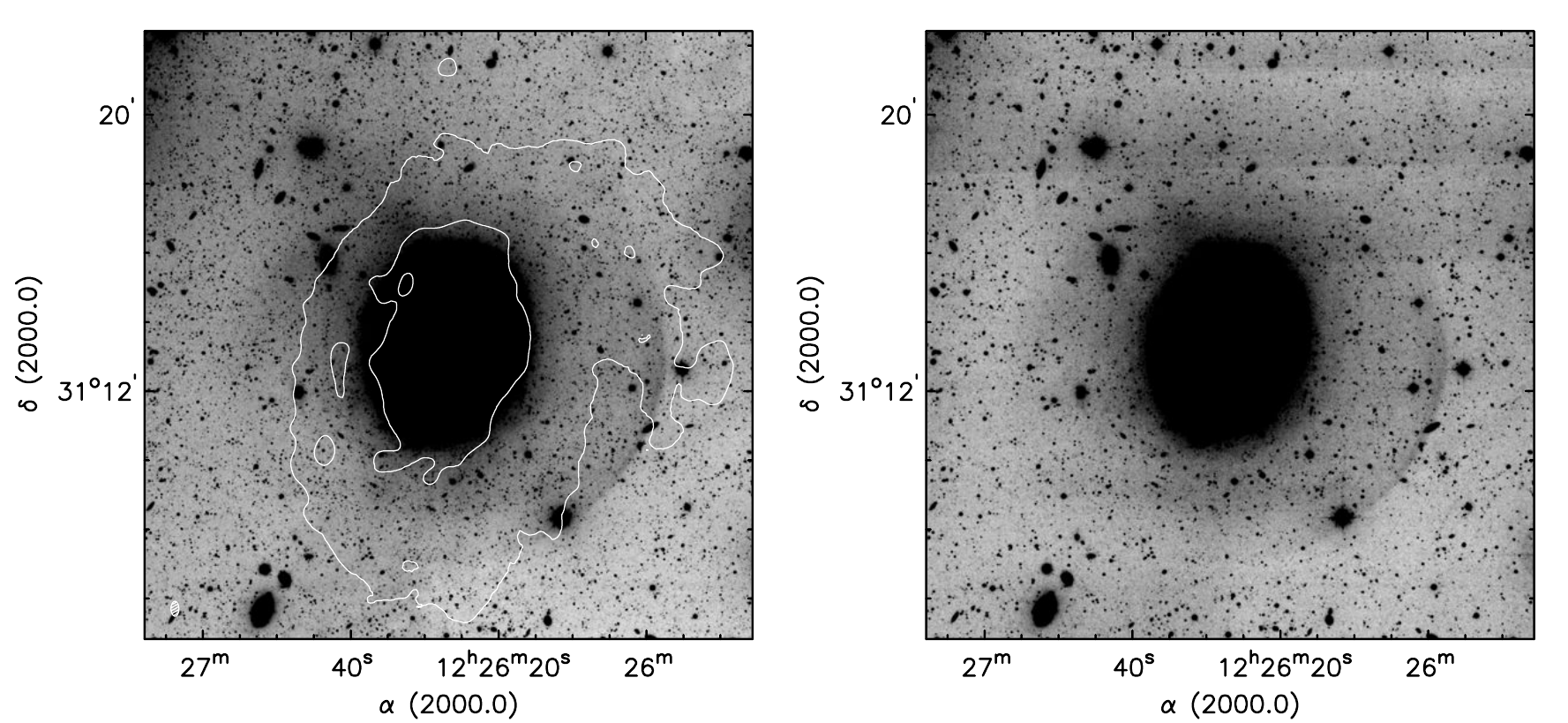}
\caption{Left: $B$-band KPNO image in grayscale with \hi\ contours
  overlaid in white. Contours shown are $(2, 20) \cdot 10^{19}$
  cm$^{-2}$. Right: the HALOSTARS $R$-band image showing similar
  features as the KPNO image.
\label{fig:annot}}
\end{figure*}

A high-contrast version of the $B$-band image is shown in
Fig.~\ref{fig:sdss} with a SDSS image of the main disk (at the same
spatial scale) superimposed.  The deep image has been smoothed to
$2''$ from its original resolution of $\sim 1.2''$ to enhance the low
surface brightness features.

We see that the main stellar disk extends much farther out than the
shallow SDSS image suggests.  This optical disk is surrounded by a low
surface brightness shell, which is most clearly defined towards the
west and SW, where we see a sharp and well-defined edge to the shell. A
faint condensation in the shell is visible to the NNE of the main disk
(at a position angle of $\sim 30^{\circ}$), marking the top of a short
``plume'' extending radially away from the disk. Stacking of
SDSS images, using the technique described in \citet{arpad} also
reveals the presence of these features (A.\ Miskoliczi, priv.\ comm.).

In the $2''$ KPNO image, the surface brightness of the shell has a
level of $\sim 3-4$ times the RMS of the background, which gives the
shell a surface brightness of $\sim 26.4$ $B$-mag arcsec$^{-2}$.
There is no clear evidence of the presence of  multiple shells.

Figure \ref{fig:annot} shows an overlay of the \hi\ on top of the
optical image. While the shell and the \hi\ disk largely coincide, we
do not see any small-scale agreement between gas and stellar
features. The shell and plume typically extend to larger radii than
the \hi\ disk.

The presence of the shell suggests NGC 4414 has undergone a minor
interaction.  Numerical simulations of interactions between a low mass
disk galaxy, and a more massive elliptical galaxy showed that the
interaction could cause shells to form \citep{quinn84,hernquist1}.
Some of these models also show the presence of a plume, as we observe
here, which is a remnant of the interloper galaxy. Later work showed
that shells could also form as the result of the interaction of a
small disk galaxy with a more massive \emph{disk} galaxy
\citep{hernquist2}, and shells have indeed been observed around some
of them \citep{schweizer}.  Recently, \citet{cooper} suggested that
the formation of shells around galaxies also occurs naturally as a
product of continuous accretion of clumps of dark matter and stars in
a cold dark matter universe.

The simulations by \citet{hernquist1} and \citet{hernquist2} show that
shell formation can occur on timescales of $\sim 10^8$ years and that
shells can persist for a few Gyr. These timescales are longer than in the
ram pressure scenario discussed above, and it is therefore likely
that the shell formation occured in a separate event.

The simulations mentioned above assumed a mass ratio of 1:100
for the interactions, but the much lower velocity dispersions of disk
galaxies compared to ellipticals should allow encounters with much
larger mass ratios to create shells in disk galaxies as well, as disk
galaxies are more easily disturbed than ellipticals
\citep{hernquist2}. A rough comparison of the luminosities of the NGC
4414 shell and extended low-surface brightness halo with that of the
main disk gives a luminosity ratio of $\sim 1:(1-5) \cdot 10^{4}$.
The asymmetry between the sharp edge of the shell towards the west and
the much more diffuse distribution in the other directions suggest an
interaction that was not entirely radial (see, e.g., Fig.~4 in
\citealt{hernquist2}). Beyond this it is difficult to derive much more
on the geometry and nature of the interaction.

Even assuming a mass ratio of 1:100 gives the intruding galaxy an
\hi\ mass of only $\sim 10^7\ M_{\odot}$ (assuming a similar $M_{\rm
  HI}/M_{\rm dyn}$ as NGC 4414). It is therefore unlikely that a
significant fraction of the gas in NGC 4414 originally belonged to the
intruder galaxy.  There are no ``smoking gun'' candidate intruder
galaxies in the immediate vicinity of NGC 4414. If the incoming galaxy
has not been completely disrupted, then one tentative candidate could
be a small galaxy, SDSS J122646.27+311904.8, that lies just beyond the
tip of the plume. The galaxy is indicated in the optical image
presented in Fig.\ \ref{fig:moms}. It distinguishes itself from other
(background) galaxies in the vicinity by being the only galaxy that is
not centrally concentrated, is more fuzzy, and has a lower surface
brightness.  Assuming this galaxy is at the same distance as NGC 4414,
its luminosity would be $\sim 1200$ times fainter than NGC
4414. Unfortunately no redshift is know for this galaxy, and no
\hi\ is detected at this position.

The interaction that resulted in the optical shell most likely did not
cause the observed \hi\ morphology due to the small mass of the
intruder galaxy.  The formation of the optical shell and the observed
{\hi\ morphology} therefore are the result of two different processes
or events. This shows that even in low-density groups such as Coma I,
deep enough observations will show that galaxies are still affected by
their environment. As noted before, many of the HALOGAS targets are in
the Coma I group and a future paper will contain a fuller discussion
of the effects of the group environment.

\section{Summary}

We have presented deep \hi\ observations of NGC 4414. These show that
NGC 4414 can be characterized by a regularly rotating, symmetrical
inner disk. An extra-planar gas component is present above the inner
disk, representing $\sim 4$ per cent of the inner disk \hi\ mass (or
about 3 per cent of the total \hi\ mass).

The outer disk is disturbed and needs to be modelled assuming a
variable dynamical center position and radial motions. This can be
interpreted as a U-shaped warp.  
%Comparison with NGC 4330, a galaxy
%with similar mass in the Virgo cluster which is currently undergoing
%ram stripping, suggests that the same process is also at work in NGC
%4414. The orientation of these ram stripping features suggest NGC 4414
%is falling in towards the NGC 4631 subgroup in Coma I.

Deep optical imaging shows the presence of an extensive low surface
brightness stellar halo containing a clear shell feature surrounding
the main galaxy. Simulations indicate that such shells are
the result of minor interactions with low mass galaxies.

Finally, the difference between conclusions reached here, of NGC 4414
being a disturbed galaxy, and the previous literature concensus (based
on more shallow observations), of NGC 4414 being undisturbed, points
at the importance of obtaining deep \hi\ and optical observations in
order to properly chararacterise the evolutionary state of a galaxy.

\begin{acknowledgements}
%We thank Bernd Vollmer for providing the simulation figures and for
%useful input to the paper.
We thank the anonymous referee for the constructive comments.
WJGdB is supported by the European Commission (grant
FP7-PEOPLE-2012-CIG \#333939).  
G.G. is a postdoctoral researcher of the FWO-Vlaanderen
(Belgium). RAMW and MP acknowledge support from the National Science
Foundation through Grant AST-0908126. RJR acknowledges support from
the National Science Foundation through Grant AST-0908106.

The Isaac Newton Telescope is operated
on the island of La Palma by the Isaac Newton Group in the Spanish
Observatorio del Roque de los Muchachos of the Instituto de
Astrofísica de Canarias. The Westerbork Synthesis Radio Telescope is
operated by the ASTRON (Netherlands Foundation for Research in
Astronomy) with support from the Netherlands Foundation for Scientific
Research NWO. Funding for the SDSS and SDSS-II has been provided by
the Alfred P. Sloan Foundation, the Participating Institutions, the
National Science Foundation, the U.S. Department of Energy, the
National Aeronautics and Space Administration, the Japanese
Monbukagakusho, the Max Planck Society, and the Higher Education
Funding Council for England. The SDSS Web Site is
http://www.sdss.org/. The SDSS is managed by the Astrophysical
Research Consortium for the Participating Institutions. The
Participating Institutions are the American Museum of Natural History,
Astrophysical Institute Potsdam, University of Basel, University of
Cambridge, Case Western Reserve University, University of Chicago,
Drexel University, Fermilab, the Institute for Advanced Study, the
Japan Participation Group, Johns Hopkins University, the Joint
Institute for Nuclear Astrophysics, the Kavli Institute for Particle
Astrophysics and Cosmology, the Korean Scientist Group, the Chinese
Academy of Sciences (LAMOST), Los Alamos National Laboratory, the
Max-Planck-Institute for Astronomy (MPIA), the Max-Planck-Institute
for Astrophysics (MPA), New Mexico State University, Ohio State
University, University of Pittsburgh, University of Portsmouth,
Princeton University, the United States Naval Observatory, and the
University of Washington.
\end{acknowledgements}

\end{document}